\documentclass[%
print,
 amsmath,amssymb,
 aps,twocolumn
]{revtex4}

\usepackage{times}%此指令是令文章字体选用Palation.
\usepackage{graphicx}% Include figure files
\usepackage{dcolumn}% Align table columns on decimal point
\usepackage{bm}% bold math
\usepackage{dsfont}
\usepackage{mathrsfs}
\usepackage[landscape,papersize={297.1mm,210mm},left=1.9cm,right=1.6cm,top=2.7cm,bottom=2.8cm]{geometry}% 此指令是规定A4纸的页边距
\usepackage[                   %dvipdfm, pdflatex,pdftex这里决定运行文件的方式不同
            pdfstartview=FitH,
            colorlinks, %注释掉此项则交叉引用为彩色边框(将colorlinks和pdfborder同时注释掉)
            pdfborder=001,   %注释掉此项则交叉引用为彩色边框
            linkcolor=blue,
            anchorcolor=blue,
            citecolor=blue,
            urlcolor=blue
            ]{hyperref}
\usepackage{amsthm}
\makeatletter

\newcommand{\Rmnum}[1]{\expandafter\@slowromancap\romannumeral #1@}
\newcommand{\tabincell}[2]{\begin{tabular}{@{}#1@{}}#2\end{tabular}}
\makeatother

\begin{document}
\title{Strong quantum nonlocality without entanglement in $n$-partite system with even $n$}

\author{Huaqi Zhou$^{1,2,3}$}

\author{Ting Gao$^{1,2,3}$}
\email{gaoting@hebtu.edu.cn}

\author{Fengli Yan$^4$}
\email{flyan@hebtu.edu.cn}
\affiliation{$^1$ School of Mathematical Sciences,
 Hebei Normal University, Shijiazhuang 050024, China \\
 $^2$ Hebei Mathematics Research Center,
 Hebei Normal University, Shijiazhuang 050024, China \\
 $^3$ Hebei International Joint Research Center for Mathematics and Interdisciplinary Science,  Hebei Normal University, Shijiazhuang 050024, China \\
$^4$ College of Physics, Hebei Key Laboratory of Photophysics Research and Application, Hebei Normal University, Shijiazhuang 050024, China}

\begin{abstract}
In multipartite systems, great progress has been made recently on the study of strong quantum nonlocality without entanglement. However, the existence of  orthogonal product sets with strong quantum nonlocality in even party systems remains unknown. Here the even number is greater than four. In this paper, we successfully construct strongly nonlocal orthogonal product sets in $n$-partite systems for all even $n$, which answers the open questions given by Halder et al. [\href{https://journals.aps.org/prl/abstract/10.1103/PhysRevLett.122.040403} {Phys. Rev. Lett \textbf{122}, 040403 (2019)}] and Yuan et al. [\href{https://journals.aps.org/pra/abstract/10.1103/PhysRevA.102.042228} {Phys. Rev. A \textbf{102}, 042228 (2020)}]  for any possible even party systems. Thus, we find general construction of strongly nonlocal orthogonal product sets in space $\otimes_{i=1}^{n}\mathcal{C}^{d_{i}}$ ($n,d_{i}\geq 3$) and show that there do exist incomplete orthogonal product bases that can be strongly nonlocal in any possible $n$-partite systems for all even $n$.
Our newly constructed orthogonal product sets are asymmetric. We analyze the differences and connections between these sets and the known orthogonal product sets in odd party systems.
 In addition, we present a local state discrimination protocol for our sets by using additional entangled resource. When at least two subsystems have dimensions greater than three, the protocol consumes less entanglement than teleportation-based protocol. Strongly nonlocal set implies that the information cannot be completely accessed as long as it does not happen that all parties are together. As an application, we connect our sets with local information hiding in multipartite system.
~\\


\end{abstract}

\maketitle

\section{Introduction}
Quantum entanglement is one of the most intriguing feature of quantum mechanics. Entangled states provide strong evidence for the validity of quantum mechanics \cite{Einstein,Werner,Horodecki,GYvE} and can be used for entanglement-assisted communication \cite{BennettFS,Buhrman,Deng,GaoYW,GYW}, quantum teleportation \cite{BennettBCJPW,GaoYL,GaoYanW} and quantum key distribution \cite{Lo,Ekert,BennettBM,YZhou} and so on. When entangled pure states violate the Bell-type inequalities, they exhibit quantum nonlocality \cite{Bell,Clauser,Freedman,Yan,Meng,Chen,DingHYG,DHYG} which means that spatially separated systems may behave in a way that cannot be explained by any local theory. This is an extremely striking quantum nonlocality, Bell nonlocality. Interestingly, there also exist other types of nonlocal behaviors, which are no longer restricted only to entangled systems \cite{BennettDFMRSSW,Niset}. When a set of orthogonal quantum states cannot be distinguished by local operations and classical communication (LOCC), it reflects the fundamental feature of quantum mechanics which is also called nonlocality \cite{BennettDFMRSSW,Niset,ZhangZGWO}. A elementary result in this area was presented by Bennett et al. who first showed a LOCC indistinguishable orthogonal product basis in $\mathcal{C}^{3}\otimes \mathcal{C}^{3}$ \cite{BennettDFMRSSW}. Inspired by this senior work of Bennett et al. on quantum nonlocality without entanglement, many orthogonal product sets (OPSs) with quantum nonlocality were provided in bipartite and multipartite systems \cite{Niset,ZhangGTCW,WangLZF,WLZF,Feng,Xu,ZhangZGWO,SHalder,Jiang,Rout21,Li21}. Among them, Rout et al. \cite{Rout21} and Li et al. \cite{Li21} studied the local indistinguishability of multiparty OPS in any possible bipartition. These results also showed that the local indistinguishability has meaningful research value for quantum data hiding \cite{Terhal,DiVincenzo,Eggeling} and quantum secret sharing \cite{Hillery,Guo,Hsu,Markham,Rahaman,JWang}.

In 2019, Halder et al. \cite{Halder} came up with the concept of strong quantum nonlocality without entanglement and showed the phenomenon by presenting two explicit strongly nonlocal orthogonal product bases on $\mathcal{C}^{3}\otimes \mathcal{C}^{3}\otimes \mathcal{C}^{3}$ and $\mathcal{C}^{4}\otimes \mathcal{C}^{4}\otimes \mathcal{C}^{4}$, respectively. A set of multiparty orthogonal product states is strongly nonlocal if it is locally irreducible in every bipartition. The locally irreducibility of multiparty orthogonal set means that it is not possible to locally eliminate one or more states from the set while preserving orthogonality of the postmeasurement states \cite{Halder}. Actually, local irreducibility is a stronger version of local indistinguishability. Because a locally irreducible set is locally indistinguishable and the inverse does not hold in general. Halder et al. \cite{Halder} left some open questions, one of which is ``whether incomplete orthogonal product bases can be strongly nonlocal".

A positive operator-valued measure (POVM) is trivial means that all the POVM elements are proportional to the identity operator \cite{Walgate}. In general, one uses the triviality of orthogonality-preserving POVM to deduce the local irreducibility and then obtains the quantum nonlocality \cite{Halder}. Recently, the study on strong quantum nonlocality has made great progress \cite{ZhangZ,Rout,Yuan,ShiLHCYWZ,ShiYCZ,Che,He,ZGY}. Yuan et al. \cite{Yuan} presented two strongly nonlocal OPSs in $\mathcal{C}^{d}\otimes \mathcal{C}^{d}\otimes \mathcal{C}^{d}$ and $\mathcal{C}^{d}\otimes \mathcal{C}^{d}\otimes \mathcal{C}^{d+1}$ and gave two explicit forms of strongly nonlocal orthogonal product basis on $\mathcal{C}^{3}\otimes \mathcal{C}^{3}\otimes \mathcal{C}^{3}\otimes \mathcal{C}^{3}$ and $\mathcal{C}^{4}\otimes \mathcal{C}^{4}\otimes \mathcal{C}^{4}\otimes \mathcal{C}^{4}$. Zhou et al. \cite{ZGY} proposed a strongly nonlocal OPS containing fewer quantum states in $\mathcal{C}^{d_{A}}\otimes \mathcal{C}^{d_{B}}\otimes \mathcal{C}^{d_{C}}$ $(d_{A},d_{B},d_{C}\geq 4)$ and generalized the structures of known OPSs to any possible three and four-partite systems. Che et al. \cite{Che} constructed strongly nonlocal unextendible product bases on quantum system $\mathcal{C}^{d_{A}}\otimes \mathcal{C}^{d_{B}}\otimes \mathcal{C}^{d_{C}}$ $(d_{A},d_{B},d_{C}\geq 3)$. In general $n$-partite systems, several strongly nonlocal sets of orthogonal entangled states were showed by Shi et al. \cite{ShiYCZ}. For odd $n$, He et al. \cite{He} presented a strongly nonlocal OPS in system $\otimes_{i=1}^{n}\mathcal{C}^{d_{i}}$ ($n,d_{i}\geq 3$). It is a partial answer to one open question in Ref. \cite{Yuan}, ``Is there any general construction of strongly nonlocal OPS in space $\otimes_{i=1}^{n}\mathcal{C}^{d_{i}}$ ($n,d_{i}\geq 3$)?''. However, the existence and general construction of strongly nonlocal OPS in $n$-partite systems for all even $n$ remains unknown.

For the strongly nonlocal OPS, another interesting phenomenon is that its indistinguishability can be changed by sharing additional entangled resource among the parties \cite{Cohen08,Rout,ZhangWZ,LiGZW,ZhangSSGQW,Bandyopadhyay18,ZGY}. One of the most general examples is to teleport \cite{BennettBCJPW} the full multipartite states to one of the parties by using enough entanglement resource and LOCC, and then to determine these states by performing appropriate measurement \cite{Bandyopadhyay16}. However, entanglement is a very valuable resource, so it is always desirable to discrimination protocols consuming less entanglement. In 2008, Cohen first proposed such protocols for certain classes of locally indistinguishable unextendible product bases \cite{Cohen08}. These protocols consume less entanglement in comparison to the teleportation-based protocols, in other word, they are resource efficient. This method motivates further research in establishing efficient quantum state local identification protocols \cite{Rout,ZhangWZ,LiGZW,ZhangSSGQW,Bandyopadhyay18,ZGY}.

In this paper, we mainly consider the OPS with strong quantum nonlocality in multipartite system. In Sec. \ref{Q2}, according to the structure given by He et al. \cite{He}, we present strongly nonlocal OPSs in any possible $n$-partite ($n$ is even) systems. These results, combined with \cite{He}, answer a question in \cite{Halder} and give a positive answer to the open question in Ref. \cite{Yuan}. Moreover, we also analyze the differences and relations between all the OPSs. In Sec. \ref{Q3}, we give an entanglement-assisted discrimination protocol to locally distinguish our OPSs. Then, we connect our OPSs with quantum data hiding in Sec. \ref{Q4}. Finally, we conclude with a brief summary in Sec. \ref{Q5}. For simplicity, all product states are not normalized.

\section{The OPS with strong nonlocality in $n$-qudit ($n$ is even) system}\label{Q2}
According to the method of He et al. \cite{He}, we first construct a set of product states in the quantum system $\otimes_{i=1}^{n}\mathcal{C}^{3}$, where $n$ is even.

Let  $|\eta_{\pm}\rangle_{i}:=|0\rangle_{i}\pm|1\rangle_{i}=|0\pm 1\rangle_{i}$ and $|\xi_{\pm}\rangle_{i}:=|1\rangle_{i}\pm|2\rangle_{i}=|1\pm2\rangle_{i}$ be the quantum states of the $i$th subsystem for $i\in I_{n}$, where $\{|0\rangle, |1\rangle, |2\rangle \}$ is the standard basis of $\mathcal{C}^{3}$  and $I_{n}$ is the index set $\{1,2,\ldots,n\}$.
Suppose that $Q$ is a subset of $I_{n}$ and $|Q|$ is odd. For any fixed $Q$, there are two subsets
\begin{equation}
\begin{aligned}
&\mathcal{C}_{Q}:=\{\mathcal{C}_{Q}^{(1)}\otimes \mathcal{C}_{Q}^{(2)}\otimes\cdots\otimes\mathcal{C}_{Q}^{(n)}\},\\
&\mathcal{D}_{Q}:=\{\mathcal{D}_{Q}^{(1)}\otimes \mathcal{D}_{Q}^{(2)}\otimes\cdots\otimes\mathcal{D}_{Q}^{(n)}\}.
\end{aligned}
\end{equation}
Here
\begin{equation}
\begin{aligned}
\mathcal{C}_{Q}^{(1)}:=\left\{\begin{aligned} &|0\rangle_{1}, &\textup{if}~1\notin Q,\\  &|\eta_{\pm}\rangle_{1}, &\textup{if}~1\in Q,\end{aligned}\right.
~~
\mathcal{D}_{Q}^{(1)}:=\left\{\begin{aligned} &|2\rangle_{1}, &\textup{if}~1\notin Q,\\  &|\xi_{\pm}\rangle_{1}, &\textup{if}~1\in Q.\end{aligned}\right.
\end{aligned}
\end{equation}
For $2\leq i\leq n$, $\mathcal{C}_{Q}^{(i)}$ and $\mathcal{D}_{Q}^{(i)}$ are defined as Table \ref{30}. Naturally, we have the set
\begin{equation}
\begin{aligned}
\mathcal{E}:=\bigcup_{Q\in \Theta, \mathcal{P}\in \{\mathcal{C},\mathcal{D}\}}\mathcal{P}_{Q},
\end{aligned}
\end{equation}
where $\Theta:=\{Q\subseteq I_{n}||Q|~\textup{is}~\textup{odd}\}$. It is easy to know that $|\Theta|=2^{n-1}$. So, there are $2^{n}$ subsets $\mathcal{P}_{Q}$. Moreover, each subset $\mathcal{P}_{Q}$ contains $2^{|Q|}$ vectors. Hence $\mathcal{E}$ contains
\begin{equation}
\begin{aligned}
2\sum_{i=1}^{\frac{n}{2}}C_{n}^{2i-1}2^{2i-1}=3^{n}-1
\end{aligned}
\end{equation}
vectors. In particular, when $n=2$, $\mathcal{E}$ is the set of the eight special product states presented by Bennett et al. \cite{BennettDFMRSSW}.

\begin{table}[tbp]
\centering
\caption{The construction of $\mathcal{C}_{Q}^{(i)}$ and $\mathcal{D}_{Q}^{(i)}$ for $2\leq i\leq n$.}\label{30}
\begin{tabular}{c|c|c}
\hline \hline
$\mathcal{P}\in \{\mathcal{C},\mathcal{D}\}$ & if $i\notin Q$ & if $i\in Q$ \\ \hline
 if $\mathcal{P}_{Q}^{(i-1)}=|0\rangle_{i-1}$ or $|\eta_{\pm}\rangle_{i-1}$~ & ~$\mathcal{P}_{Q}^{(i)}=|0\rangle_{i}$~ & ~$\mathcal{P}_{Q}^{(i)}=|\xi_{\pm}\rangle_{i}$~ \\ \hline
 if $\mathcal{P}_{Q}^{(i-1)}=|2\rangle_{i-1}$ or $|\xi_{\pm}\rangle_{i-1}$~ & ~$\mathcal{P}_{Q}^{(i)}=|2\rangle_{i}$~ & ~$\mathcal{P}_{Q}^{(i)}=|\eta_{\pm}\rangle_{i}$~ \\ \hline
\end{tabular}
\end{table}

In $\otimes_{i=1}^{n+1}\mathcal{C}^{3}$, He et al. \cite{He} gave the OPS $\mathcal{O}:=\bigcup_{K\in \Lambda, \mathcal{P}\in \{\mathcal{C},\mathcal{D}\}}\mathcal{P}_{K}$ with strongly quantum nonlocality, where $n$ is even and $\Lambda:=\{K\subseteq I_{n+1}||K|~\textup{is}~\textup{even}\}$. The difference between these two sets $\mathcal{E}$ and $\mathcal{O}$ behaves in the difference of index sets $Q$ and $K$. Now, we analyze the relationship between these two sets. Unless otherwise specified, $\mathcal{E}$ and $\mathcal{O}$ refer to the sets in $n$ and $(n+1)$-qutrit quantum systems, respectively.

It is known that set $\mathcal{O}$ is symmetric. That is, it is invariant under cyclic permutation of the parties. Without loss of generality, we consider removing the $(n+1)$th subsystem of set $\mathcal{O}$. We first remove the quantum states with $|0\rangle_{n+1}$ and $|2\rangle_{n+1}$ from $\mathcal{O}$ and mark the set of the remaining quantum states as $\mathcal{E}_{1}$. Then we remove the particles of the $(n+1)$th subsystem in the set $\mathcal{E}_{1}$. The resulting quantum states form a new set $\mathcal{E}_{2}$. Interestingly, this new set happens to be the set $\mathcal{E}$. Moreover, if one repeats the above two steps to remove the $n$th subsystem of the set $\mathcal{E}$, one can get the set $\mathcal{O}$ in $(n-1)$-qutrit quantum system. For the detailed explanation please refer to Appendix \ref{A}. To help ones understand even better, we give a simple example to illustrate.

\emph{Example 1}. In $5$-qutrit quantum system, the set $\mathcal{O}$ is described as
\begin{equation*}
\begin{aligned}
&\mathcal{C}_{K_{1}}:=\{|0\rangle_{1}|0\rangle_{2}|0\rangle_{3}|0\rangle_{4}|0\rangle_{5}\},\\
&\mathcal{C}_{K_{2}}:=\{|\eta_{\pm}\rangle_{1}|\xi_{\pm}\rangle_{2}|2\rangle_{3}|2\rangle_{4}|2\rangle_{5}\},\\
&\mathcal{C}_{K_{3}}:=\{|\eta_{\pm}\rangle_{1}|0\rangle_{2}|\xi_{\pm}\rangle_{3}|2\rangle_{4}|2\rangle_{5}\},\\
&\mathcal{C}_{K_{4}}:=\{|\eta_{\pm}\rangle_{1}|0\rangle_{2}|0\rangle_{3}|\xi_{\pm}\rangle_{4}|2\rangle_{5}\},\\
&\mathcal{C}_{K_{5}}:=\{|\eta_{\pm}\rangle_{1}|0\rangle_{2}|0\rangle_{3}|0\rangle_{4}|\xi_{\pm}\rangle_{5}\},\\
&\mathcal{C}_{K_{6}}:=\{|0\rangle_{1}|\xi_{\pm}\rangle_{2}|\eta_{\pm}\rangle_{3}|0\rangle_{4}|0\rangle_{5}\},\\
&\mathcal{C}_{K_{7}}:=\{|0\rangle_{1}|\xi_{\pm}\rangle_{2}|2\rangle_{3}|\eta_{\pm}\rangle_{4}|0\rangle_{5}\},\\
&\mathcal{C}_{K_{8}}:=\{|0\rangle_{1}|\xi_{\pm}\rangle_{2}|2\rangle_{3}|2\rangle_{4}|\eta_{\pm}\rangle_{5}\},\\
&\mathcal{C}_{K_{9}}:=\{|0\rangle_{1}|0\rangle_{2}|\xi_{\pm}\rangle_{3}|\eta_{\pm}\rangle_{4}|0\rangle_{5}\},\\
&\mathcal{C}_{K_{10}}:=\{|0\rangle_{1}|0\rangle_{2}|\xi_{\pm}\rangle_{3}|2\rangle_{4}|\eta_{\pm}\rangle_{5}\},\\
&\mathcal{C}_{K_{11}}:=\{|0\rangle_{1}|0\rangle_{2}|0\rangle_{3}|\xi_{\pm}\rangle_{4}|\eta_{\pm}\rangle_{5}\},\\
\end{aligned}
\end{equation*}
\begin{equation*}
\begin{aligned}
&\mathcal{C}_{K_{12}}:=\{|\eta_{\pm}\rangle_{1}|\xi_{\pm}\rangle_{2}|\eta_{\pm}\rangle_{3}|\xi_{\pm}\rangle_{4}|2\rangle_{5}\},\\
&\mathcal{C}_{K_{13}}:=\{|\eta_{\pm}\rangle_{1}|\xi_{\pm}\rangle_{2}|\eta_{\pm}\rangle_{3}|0\rangle_{4}|\xi_{\pm}\rangle_{5}\},\\
&\mathcal{C}_{K_{14}}:=\{|\eta_{\pm}\rangle_{1}|\xi_{\pm}\rangle_{2}|2\rangle_{3}|\eta_{\pm}\rangle_{4}|\xi_{\pm}\rangle_{5}\},\\
&\mathcal{C}_{K_{15}}:=\{|\eta_{\pm}\rangle_{1}|0\rangle_{2}|\xi_{\pm}\rangle_{3}|\eta_{\pm}\rangle_{4}|\xi_{\pm}\rangle_{5}\},\\
&\mathcal{C}_{K_{16}}:=\{|0\rangle_{1}|\xi_{\pm}\rangle_{2}|\eta_{\pm}\rangle_{3}|\xi_{\pm}\rangle_{4}|\eta_{\pm}\rangle_{5}\},\\
&\mathcal{D}_{K_{1}}:=\{|2\rangle_{1}|2\rangle_{2}|2\rangle_{3}|2\rangle_{4}|2\rangle_{5}\},\\
&\mathcal{D}_{K_{2}}:=\{|\xi_{\pm}\rangle_{1}|\eta_{\pm}\rangle_{2}|0\rangle_{3}|0\rangle_{4}|0\rangle_{5}\},\\
&\mathcal{D}_{K_{3}}:=\{|\xi_{\pm}\rangle_{1}|2\rangle_{2}|\eta_{\pm}\rangle_{3}|0\rangle_{4}|0\rangle_{5}\},\\
&\mathcal{D}_{K_{4}}:=\{|\xi_{\pm}\rangle_{1}|2\rangle_{2}|2\rangle_{3}|\eta_{\pm}\rangle_{4}|0\rangle_{5}\},\\
&\mathcal{D}_{K_{5}}:=\{|\xi_{\pm}\rangle_{1}|2\rangle_{2}|2\rangle_{3}|2\rangle_{4}|\eta_{\pm}\rangle_{5}\},\\
&\mathcal{D}_{K_{6}}:=\{|2\rangle_{1}|\eta_{\pm}\rangle_{2}|\xi_{\pm}\rangle_{3}|2\rangle_{4}|2\rangle_{5}\},\\
&\mathcal{D}_{K_{7}}:=\{|2\rangle_{1}|\eta_{\pm}\rangle_{2}|0\rangle_{3}|\xi_{\pm}\rangle_{4}|2\rangle_{5}\},\\
&\mathcal{D}_{K_{8}}:=\{|2\rangle_{1}|\eta_{\pm}\rangle_{2}|0\rangle_{3}|0\rangle_{4}|\xi_{\pm}\rangle_{5}\},\\
&\mathcal{D}_{K_{9}}:=\{|2\rangle_{1}|2\rangle_{2}|\eta_{\pm}\rangle_{3}|\xi_{\pm}\rangle_{4}|2\rangle_{5}\},\\
&\mathcal{D}_{K_{10}}:=\{|2\rangle_{1}|2\rangle_{2}|\eta_{\pm}\rangle_{3}|0\rangle_{4}|\xi_{\pm}\rangle_{5}\},\\
&\mathcal{D}_{K_{11}}:=\{|2\rangle_{1}|2\rangle_{2}|2\rangle_{3}|\eta_{\pm}\rangle_{4}|\xi_{\pm}\rangle_{5}\},\\
&\mathcal{D}_{K_{12}}:=\{|\xi_{\pm}\rangle_{1}|\eta_{\pm}\rangle_{2}|\xi_{\pm}\rangle_{3}|\eta_{\pm}\rangle_{4}|0\rangle_{5}\},\\
&\mathcal{D}_{K_{13}}:=\{|\xi_{\pm}\rangle_{1}|\eta_{\pm}\rangle_{2}|\xi_{\pm}\rangle_{3}|2\rangle_{4}|\eta_{\pm}\rangle_{5}\},\\
&\mathcal{D}_{K_{14}}:=\{|\xi_{\pm}\rangle_{1}|\eta_{\pm}\rangle_{2}|0\rangle_{3}|\xi_{\pm}\rangle_{4}|\eta_{\pm}\rangle_{5}\},\\
&\mathcal{D}_{K_{15}}:=\{|\xi_{\pm}\rangle_{1}|2\rangle_{2}|\eta_{\pm}\rangle_{3}|\xi_{\pm}\rangle_{4}|\eta_{\pm}\rangle_{5}\},\\
&\mathcal{D}_{K_{16}}:=\{|2\rangle_{1}|\eta_{\pm}\rangle_{2}|\xi_{\pm}\rangle_{3}|\eta_{\pm}\rangle_{4}|\xi_{\pm}\rangle_{5}\}.\\
\end{aligned}
\end{equation*}
Here,
\begin{equation*}
\begin{aligned}
&K_{1}=\emptyset, && K_{9}=\{3,4\}, \\
&K_{2}=\{1,2\}, && K_{10}=\{3,5\}, \\
&K_{3}=\{1,3\}, && K_{11}=\{4,5\}, \\
&K_{4}=\{1,4\}, && K_{12}=\{1,2,3,4\}, \\
&K_{5}=\{1,5\}, && K_{13}=\{1,2,3,5\}, \\
&K_{6}=\{2,3\}, && K_{14}=\{1,2,4,5\}, \\
&K_{7}=\{2,4\}, && K_{15}=\{1,3,4,5\}, \\
&K_{8}=\{2,5\}, && K_{16}=\{2,3,4,5\}. \\
\end{aligned}
\end{equation*}
We consider removing the $5$th subsystem.

Step 1. Removing the quantum states with $|0\rangle_{5}$ and $|2\rangle_{5}$, actually, this is to remove the subsets whose corresponding index sets do not contain element $5$. Then, the remaining subsets are as following
\begin{equation*}
\begin{aligned}
&\mathcal{C}_{K_{5}}:=\{|\eta_{\pm}\rangle_{1}|0\rangle_{2}|0\rangle_{3}|0\rangle_{4}|\xi_{\pm}\rangle_{5}\},\\
&\mathcal{C}_{K_{8}}:=\{|0\rangle_{1}|\xi_{\pm}\rangle_{2}|2\rangle_{3}|2\rangle_{4}|\eta_{\pm}\rangle_{5}\},\\
&\mathcal{C}_{K_{10}}:=\{|0\rangle_{1}|0\rangle_{2}|\xi_{\pm}\rangle_{3}|2\rangle_{4}|\eta_{\pm}\rangle_{5}\},\\
&\mathcal{C}_{K_{11}}:=\{|0\rangle_{1}|0\rangle_{2}|0\rangle_{3}|\xi_{\pm}\rangle_{4}|\eta_{\pm}\rangle_{5}\},\\
&\mathcal{C}_{K_{13}}:=\{|\eta_{\pm}\rangle_{1}|\xi_{\pm}\rangle_{2}|\eta_{\pm}\rangle_{3}|0\rangle_{4}|\xi_{\pm}\rangle_{5}\},\\
&\mathcal{C}_{K_{14}}:=\{|\eta_{\pm}\rangle_{1}|\xi_{\pm}\rangle_{2}|2\rangle_{3}|\eta_{\pm}\rangle_{4}|\xi_{\pm}\rangle_{5}\},\\
&\mathcal{C}_{K_{15}}:=\{|\eta_{\pm}\rangle_{1}|0\rangle_{2}|\xi_{\pm}\rangle_{3}|\eta_{\pm}\rangle_{4}|\xi_{\pm}\rangle_{5}\},\\
&\mathcal{C}_{K_{16}}:=\{|0\rangle_{1}|\xi_{\pm}\rangle_{2}|\eta_{\pm}\rangle_{3}|\xi_{\pm}\rangle_{4}|\eta_{\pm}\rangle_{5}\},\\
\end{aligned}
\end{equation*}
\begin{equation*}
\begin{aligned}
&\mathcal{D}_{K_{5}}:=\{|\xi_{\pm}\rangle_{1}|2\rangle_{2}|2\rangle_{3}|2\rangle_{4}|\eta_{\pm}\rangle_{5}\},\\
&\mathcal{D}_{K_{8}}:=\{|2\rangle_{1}|\eta_{\pm}\rangle_{2}|0\rangle_{3}|0\rangle_{4}|\xi_{\pm}\rangle_{5}\},\\
&\mathcal{D}_{K_{10}}:=\{|2\rangle_{1}|2\rangle_{2}|\eta_{\pm}\rangle_{3}|0\rangle_{4}|\xi_{\pm}\rangle_{5}\},\\
&\mathcal{D}_{K_{11}}:=\{|2\rangle_{1}|2\rangle_{2}|2\rangle_{3}|\eta_{\pm}\rangle_{4}|\xi_{\pm}\rangle_{5}\},\\
&\mathcal{D}_{K_{13}}:=\{|\xi_{\pm}\rangle_{1}|\eta_{\pm}\rangle_{2}|\xi_{\pm}\rangle_{3}|2\rangle_{4}|\eta_{\pm}\rangle_{5}\},\\
&\mathcal{D}_{K_{14}}:=\{|\xi_{\pm}\rangle_{1}|\eta_{\pm}\rangle_{2}|0\rangle_{3}|\xi_{\pm}\rangle_{4}|\eta_{\pm}\rangle_{5}\},\\
&\mathcal{D}_{K_{15}}:=\{|\xi_{\pm}\rangle_{1}|2\rangle_{2}|\eta_{\pm}\rangle_{3}|\xi_{\pm}\rangle_{4}|\eta_{\pm}\rangle_{5}\},\\
&\mathcal{D}_{K_{16}}:=\{|2\rangle_{1}|\eta_{\pm}\rangle_{2}|\xi_{\pm}\rangle_{3}|\eta_{\pm}\rangle_{4}|\xi_{\pm}\rangle_{5}\}.\\
\end{aligned}
\end{equation*}
Their union is the set $\mathcal{E}_{1}$. It is a subset of the set $\mathcal{O}$ in $5$-qutrit quantum system.

Step 2. We remove the particles of the $5$th subsystem in set $\mathcal{E}_{1}$ and get the set $\mathcal{E}_{2}$. It is given by
\begin{equation*}
\begin{aligned}
&\mathcal{C}_{K_{5}'}:=\{|\eta_{\pm}\rangle_{1}|0\rangle_{2}|0\rangle_{3}|0\rangle_{4}\},\\
&\mathcal{C}_{K_{8}'}:=\{|0\rangle_{1}|\xi_{\pm}\rangle_{2}|2\rangle_{3}|2\rangle_{4}\},\\
&\mathcal{C}_{K_{10}'}:=\{|0\rangle_{1}|0\rangle_{2}|\xi_{\pm}\rangle_{3}|2\rangle_{4}\},\\
&\mathcal{C}_{K_{11}'}:=\{|0\rangle_{1}|0\rangle_{2}|0\rangle_{3}|\xi_{\pm}\rangle_{4}\},\\
&\mathcal{C}_{K_{13}'}:=\{|\eta_{\pm}\rangle_{1}|\xi_{\pm}\rangle_{2}|\eta_{\pm}\rangle_{3}|0\rangle_{4}\},\\
&\mathcal{C}_{K_{14}'}:=\{|\eta_{\pm}\rangle_{1}|\xi_{\pm}\rangle_{2}|2\rangle_{3}|\eta_{\pm}\rangle_{4}\},\\
&\mathcal{C}_{K_{15}'}:=\{|\eta_{\pm}\rangle_{1}|0\rangle_{2}|\xi_{\pm}\rangle_{3}|\eta_{\pm}\rangle_{4}\},\\
&\mathcal{C}_{K_{16}'}:=\{|0\rangle_{1}|\xi_{\pm}\rangle_{2}|\eta_{\pm}\rangle_{3}|\xi_{\pm}\rangle_{4}\},\\
&\mathcal{D}_{K_{5}'}:=\{|\xi_{\pm}\rangle_{1}|2\rangle_{2}|2\rangle_{3}|2\rangle_{4}\},\\
&\mathcal{D}_{K_{8}'}:=\{|2\rangle_{1}|\eta_{\pm}\rangle_{2}|0\rangle_{3}|0\rangle_{4}\},\\
&\mathcal{D}_{K_{10}'}:=\{|2\rangle_{1}|2\rangle_{2}|\eta_{\pm}\rangle_{3}|0\rangle_{4}\},\\
&\mathcal{D}_{K_{11}'}:=\{|2\rangle_{1}|2\rangle_{2}|2\rangle_{3}|\eta_{\pm}\rangle_{4}\},\\
&\mathcal{D}_{K_{13}'}:=\{|\xi_{\pm}\rangle_{1}|\eta_{\pm}\rangle_{2}|\xi_{\pm}\rangle_{3}|2\rangle_{4}\},\\
&\mathcal{D}_{K_{14}'}:=\{|\xi_{\pm}\rangle_{1}|\eta_{\pm}\rangle_{2}|0\rangle_{3}|\xi_{\pm}\rangle_{4}\},\\
&\mathcal{D}_{K_{15}'}:=\{|\xi_{\pm}\rangle_{1}|2\rangle_{2}|\eta_{\pm}\rangle_{3}|\xi_{\pm}\rangle_{4}\},\\
&\mathcal{D}_{K_{16}'}:=\{|2\rangle_{1}|\eta_{\pm}\rangle_{2}|\xi_{\pm}\rangle_{3}|\eta_{\pm}\rangle_{4}\}.\\
\end{aligned}
\end{equation*}
Meanwhile, the index sets become as follows
\begin{equation*}
\begin{aligned}
&K_{5}'=\{1\}, && K_{13}'=\{1,2,3\}, \\
&K_{8}'=\{2\}, && K_{14}'=\{1,2,4\}, \\
&K_{10}'=\{3\}, && K_{15}'=\{1,3,4\}, \\
&K_{11}'=\{4\}, && K_{16}'=\{2,3,4\}. \\
\end{aligned}
\end{equation*}
We find that the collection $\{K_{5}',K_{8}',K_{10}',K_{11}',K_{13}',K_{14}',$ $K_{15}',K_{16}'\}$ of these index sets happens to be the index collection $\Theta$ in $4$-qutrit quantum system. So, the set $\mathcal{E}_{2}$ is equal to the set $\mathcal{E}$ in $4$-qutrit quantum system.

Repeating above two steps to remove the $4$th subsystem of set $\mathcal{E}$ in $4$-qutrit quantum system, we obtain the following subsets
\begin{equation*}
\begin{aligned}
&\mathcal{C}_{K_{11}''}:=\{|0\rangle_{1}|0\rangle_{2}|0\rangle_{3}\},\\
&\mathcal{C}_{K_{14}''}:=\{|\eta_{\pm}\rangle_{1}|\xi_{\pm}\rangle_{2}|2\rangle_{3}\},\\
&\mathcal{C}_{K_{15}''}:=\{|\eta_{\pm}\rangle_{1}|0\rangle_{2}|\xi_{\pm}\rangle_{3}\},\\
&\mathcal{C}_{K_{16}''}:=\{|0\rangle_{1}|\xi_{\pm}\rangle_{2}|\eta_{\pm}\rangle_{3}\},\\
&\mathcal{D}_{K_{11}''}:=\{|2\rangle_{1}|2\rangle_{2}|2\rangle_{3}\},\\
&\mathcal{D}_{K_{14}''}:=\{|\xi_{\pm}\rangle_{1}|\eta_{\pm}\rangle_{2}|0\rangle_{3}\},\\
&\mathcal{D}_{K_{15}''}:=\{|\xi_{\pm}\rangle_{1}|2\rangle_{2}|\eta_{\pm}\rangle_{3}\},\\
&\mathcal{D}_{K_{16}''}:=\{|2\rangle_{1}|\eta_{\pm}\rangle_{2}|\xi_{\pm}\rangle_{3}\},\\
\end{aligned}
\end{equation*}
where,
\begin{equation*}
\begin{aligned}
&K_{11}''=\emptyset, && K_{15}''=\{1,3\}, \\
&K_{14}''=\{1,2\}, && K_{16}''=\{2,3\}. \\
\end{aligned}
\end{equation*}
Obviously, set $\mathcal{O}$ in $3$-qutrit quantum system is the union of these new subsets.
\hfill $\square$

It is worth noting that the set $\mathcal{E}$ is asymmetric and each of its cyclic permutations under the parties can be obtained by removing different subsystem of the set $\mathcal{O}$. Additionally, by removing the $i$th $(i\in I_{n})$ subsystem of set $\mathcal{E}$, we can get a new set $\mathcal{O}_{2}$. It will become the set $\mathcal{O}$ in $(n-1)$-qutrit quantum system, when we transform $|0\rangle_{j}$, $|\eta_{\pm}\rangle_{j}$, $|\xi_{\pm}\rangle_{j}$ and $|2\rangle_{j}$ to $|2\rangle_{j}$, $|\xi_{\pm}\rangle_{j}$, $|\eta_{\pm}\rangle_{j}$ and $|0\rangle_{j}$ for all $j>i$, respectively. Appendix \ref{B} has the detailed illustration. By using the relationship, we can obtain the following two theorems.

\emph{Theorem 1}. The set $\mathcal{E}$ is an OPS.

$Proof$. According to the set $\mathcal{O}$, we know that $\mathcal{E}_{1}$ is an OPS and is also the union of all $\mathcal{P}_{K}$ satisfying that $\mathcal{P}_{K}^{(n+1)}$ is equal to $|\eta_{\pm}\rangle_{n+1}$ or $|\xi_{\pm}\rangle_{n+1}$. Suppose $\Lambda_{1}$ is the collection of all index sets $K$ with element $n+1$, we have $\mathcal{E}_{1}=\cup_{K\in \Lambda_{1},~\mathcal{P}\in \{\mathcal{C},\mathcal{D}\}} \mathcal{P}_{K}$. Meanwhile, $\mathcal{E}=\cup_{K\in \Lambda_{1},~\mathcal{P}\in \{\mathcal{C},\mathcal{D}\}} \mathcal{P}_{K}^{(N)}$, where $\mathcal{P}_{K}^{(N)}=\{\mathcal{P}_{K}^{(1)}\otimes\cdots\otimes\mathcal{P}_{K}^{(n)}\}$.

For any two different index sets $K_{1},K_{2}\in \Lambda_{1}$, it is obvious that $\mathcal{P}_{K_{1}}$ and $\mathcal{P}_{K_{2}}$ are mutually orthogonal but $\mathcal{P}_{K_{1}}^{(n+1)}$ and $\mathcal{P}_{K_{2}}^{(n+1)}$ are not. Here we say two sets $S_{1}$ and $S_{2}$ are mutually orthogonal if $_{1}\langle\varphi|\varphi\rangle_{2}= 0$ for any two states $|\varphi\rangle_{1}\in S_{1}$ and $|\varphi\rangle_{2}\in S_{2}$. Then, we can deduce that $\mathcal{P}_{K_{1}}^{(N)}$ and $\mathcal{P}_{K_{2}}^{(N)}$ are mutually orthogonal. In addition, because each $\mathcal{P}_{K}^{(N)}$ is an OPS. So, any two states in set $\mathcal{E}$ are mutually orthogonal. That is, the set $\mathcal{E}$ is also an OPS.
\hfill $\square$

\emph{Theorem 2}. In the quantum system $\otimes_{i=1}^{n}\mathcal{C}^{3}$ $(n$ is even and $n>3)$, the set $\mathcal{E}$ is strongly nonlocal.

Consider strong nonlocality of OPS $\mathcal{E}$, we only need to show that each party $X_{i}$ $(=I_{n}\setminus \{i\},i\in I_{n})$ can only perform a trivial orthogonality-preserving POVM \cite{ShiLHCYWZ}. This can be proved by using \cite[Theorem 1]{ZGY} because every $\mathcal{P}_{Q}$ is a basis spanned by the computational basis $\mathcal{P}_{Q}^{[1]}\otimes \mathcal{P}_{Q}^{[2]}\otimes\cdots\otimes\mathcal{P}_{Q}^{[n]}$ of the corresponding subspace and every component of the quantum state in $\mathcal{P}_{Q}$ is nonzero under this computational basis. Here $\mathcal{P}_{Q}^{[i]}$ $(i\in I_{n})$ is a subset of the computational basis of the $i$th subsystem. Specifically,
\begin{equation}\label{14}
\mathcal{P}_{Q}^{[i]}=
\left\{
\begin{aligned}
&\{|0\rangle_{i}\},&& \textup{if}~\mathcal{P}_{Q}^{(i)}=|0\rangle_{i},\\
&\{|0\rangle_{i},|1\rangle_{i}\},&&\textup{if}~\mathcal{P}_{Q}^{(i)}=|\eta\rangle_{i},\\
&\{|1\rangle_{i},|2\rangle_{i}\},&&\textup{if}~\mathcal{P}_{Q}^{(i)}=|\xi\rangle_{i},\\
&\{|2\rangle_{i}\},&&\textup{if}~\mathcal{P}_{Q}^{(i)}=|2\rangle_{i},
\end{aligned}
\right.
\end{equation}
for $i\in I_{n}$.

\begin{figure}[h]
\centering
\includegraphics[width=0.48\textwidth]{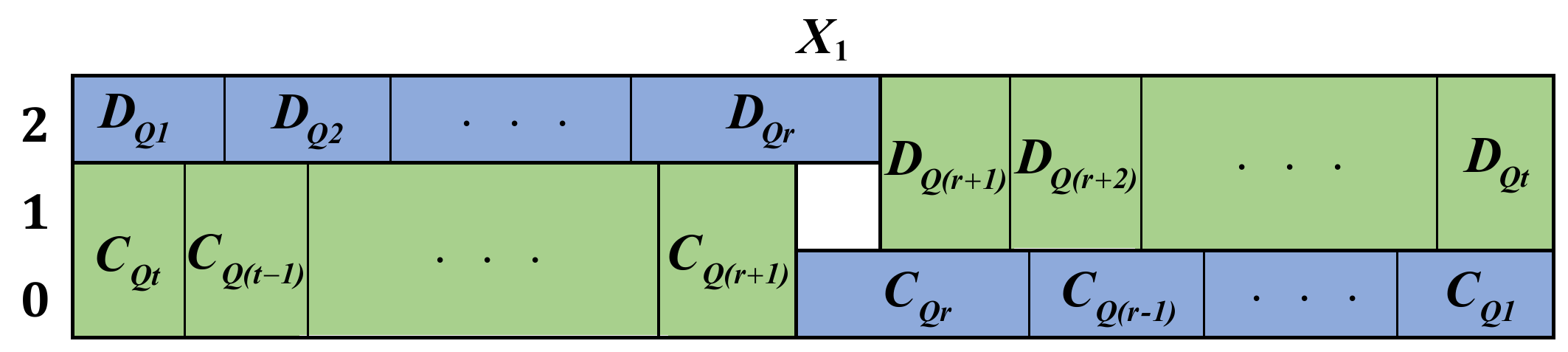}
\caption{This is the plane tile of set $\mathcal{E}$ in $1|X_{1}$ bipartition. Each subset $\mathcal{P}_{Q}$ of $\mathcal{E}$ corresponds a domain in this diagram. Actually, the projection set of subset $\mathcal{P}_{Q}$ on the $\{1\}$ (or $X_{1}$) party is the coordinate of the corresponding grid on the $\{1\}$ (or $X_{1}$) party. The projection set of every $\mathcal{C}_{Q}$ on the $\{1\}$ party is either $\{|0\rangle_{1}\}$ or $\{|0\rangle_{1},|1\rangle_{1}\}$, while the projection set of every $\mathcal{D}_{Q}$ on the $\{1\}$ party is either $\{|2\rangle_{1}\}$ or $\{|1\rangle_{1},|2\rangle_{1}\}$. Here, $Q\in\Theta=\{Q1,Q2,\ldots,Qt\}$ and $Qr=\{2,\ldots,n\}$. \label{11}}
\end{figure}

We first consider the orthogonality-preserving POVM on party $X_{1}$. Assume that $\mathcal{P}_{Q}^{[X_{1}]}=\mathcal{P}_{Q}^{[2]}\otimes \mathcal{P}_{Q}^{[3]}\otimes \cdots \otimes \mathcal{P}_{Q}^{[n]}$ represents the projection set of $\mathcal{P}_{Q}$ on the $X_{1}$ party. If a family of projection sets $\{\mathcal{P}_{Q}^{[X_{1}]}\}_{\mathcal{P}_{Q}\subset \mathcal{E}}$ satisfies
\begin{equation}\label{1}
\begin{aligned}
\left(\bigcup\limits_{\mathcal{P}_{Q}\subset \mathcal{E}'}\mathcal{P}_{Q}^{[X_{1}]}\right)\bigcap \left(\bigcup\limits_{\mathcal{P}_{Q}\subset \mathcal{E}\setminus\mathcal{E}'}\mathcal{P}_{Q}^{[X_{1}]}\right)\neq \emptyset,
\end{aligned}
\end{equation}
for all $\mathcal{E}'$, which is the union of some $\mathcal{P}_{Q}$ and requires $\mathcal{E}'\subsetneqq \mathcal{E}$, then we call it connected, or we can say that $\{\mathcal{P}_{Q}\}_{\mathcal{P}_{Q}\subset \mathcal{E}}$ is connected on party $X_{1}$ \cite{ZGY}. Fig. \ref{11} shows the plane structure of the set $\mathcal{E}$ in $1|X_{1}$ bipartition. By observing the simple structure, it is not difficult to find that we can achieve our goal as long as the connectedness of this family of subsets $\{\mathcal{P}_{Q}\}_{\mathcal{P}_{Q}\subset \mathcal{E}}$ on party $X_{1}$ is proved. This can be associated with the strong nonlocality of $\mathcal{O}$ in system $\otimes_{i=2}^{n}\mathcal{C}^{3}$. For other $X_{i}$ $(i=\{2,\ldots,n\})$ party, we can use similar methods. The detailed proof is given in Appendix \ref{D}.

Next, by extending the dimension of the grid in Fig. \ref{11}, we can generalize the structure of the set $\mathcal{E}$ to any finite dimension. We just need to replace $|\eta_{\pm}\rangle_{t}$, $|\xi_{\pm}\rangle_{t}$ and $|2\rangle_{t}$ with $\{|\alpha_{i}\rangle_{t}\}:=\{\sum_{u=0}^{d_{t}-2}\omega_{d_{t}-1}^{iu}|u\rangle\}_{i=0}^{d_{t}-2}$, $\{|\beta_{j}\rangle_{t}\}:=\{\sum_{u=0}^{d_{t}-2}\omega_{d_{t}-1}^{ju}|u+1\rangle\}_{j=0}^{d_{t}-2}$ and $|d_{t}-1\rangle_{t}$, respectively. Here $d_{t}$ is the dimension of the $t$th subsystem, $\omega_{d_{t}-1}:=\textup{e}^{\frac{2\pi\textup{i}}{d_{t}-1}}$. For any fixed $Q$, there are two subsets
\begin{equation}
\begin{aligned}
&\mathcal{C}_{Q}^{f}:=\mathcal{C}_{Q}^{f(1)}\otimes \mathcal{C}_{Q}^{f(2)}\otimes\cdots\otimes\mathcal{C}_{Q}^{f(n)},\\
&\mathcal{D}_{Q}^{f}:=\mathcal{D}_{Q}^{f(1)}\otimes \mathcal{D}_{Q}^{f(2)}\otimes\cdots\otimes\mathcal{D}_{Q}^{f(n)}.
\end{aligned}
\end{equation}
Let
\begin{equation}
\begin{aligned}
&\mathcal{C}_{Q}^{f(1)}:=\left\{\begin{aligned} &|0\rangle_{1}, &\textup{if}~1\notin Q,\\  &\{|\alpha_{i}\rangle_{1}\}, &\textup{if}~1\in Q,\end{aligned}\right.
\\
&\mathcal{D}_{Q}^{f(1)}:=\left\{\begin{aligned} &|d_{1}-1\rangle_{1}, &\textup{if}~1\notin Q,\\  &\{|\beta_{j}\rangle_{1}\}, &\textup{if}~1\in Q.\end{aligned}\right.
\end{aligned}
\end{equation}
For $2\leq t\leq n$, $\mathcal{C}_{Q}^{f(t)}$ and $\mathcal{D}_{Q}^{f(t)}$ are defined as Table \ref{31}.

\begin{table}[tbp]
\centering
\caption{The construction of $\mathcal{C}_{Q}^{f(t)}$ and $\mathcal{D}_{Q}^{f(t)}$ for $2\leq t\leq n$.}\label{31}
\begin{tabular}{c|c|c}
 \hline \hline
$\mathcal{P}\in \{\mathcal{C},\mathcal{D}\}$ & if $t\notin Q$ & if $t\in Q$  \\ \hline
 \tabincell{l}{if $\mathcal{P}_{Q}^{f(t-1)}=|0\rangle_{t-1}$\\~~~~~~~~~~~~~~~~~ or $\{|\alpha_{i}\rangle_{t-1}\}$} & $\mathcal{P}_{Q}^{f(t)}=|0\rangle_{t}$ & $\mathcal{P}_{Q}^{f(t)}=\{|\beta_{j}\rangle_{t}\}$  \\ \hline
 \tabincell{l}{if $\mathcal{P}_{Q}^{f(t-1)}=|d_{t-1}-1\rangle_{t-1}$\\~~~~~~~~~~~~~~~~~ or $\{|\beta_{j}\rangle_{t-1}\}$} & $\mathcal{P}_{Q}^{f(t)}=|d_{t}-1\rangle_{t}$ & $\mathcal{P}_{Q}^{f(t)}=\{|\alpha_{i}\rangle_{t}\}$  \\ \hline
\end{tabular}
\end{table}

Then, we get the new set $\mathcal{F}:=\bigcup_{Q\in \Theta, \mathcal{P}\in \{\mathcal{C},\mathcal{D}\}}\mathcal{P}_{Q}^{f}$. Because the set $\mathcal{F}$ has the same structure as the set $\mathcal{E}$, the four conditions of \cite[Theorem 1]{ZGY} still hold. Therefore, we have the following theorem.

\emph{Theorem 3}. In the quantum system $\otimes_{i=1}^{n}\mathcal{C}^{d_{i}}$ $(n$ is even, $n>3$ and $d_{i}\geq 3)$, the set $\mathcal{F}$ is an OPS with strong nonlocality and
\begin{equation*}
\begin{aligned}
|\mathcal{F}|=d_{1}d_{2}\cdots d_{n}-(d_{1}-2)(d_{2}-2)\cdots (d_{n}-2).
\end{aligned}
\end{equation*}

Theorem 3 is not only a positive answer to the open question of ``whether incomplete orthogonal product bases can be strongly nonlocal'' raised by Halder et al. \cite{Halder}, but also a partial answer to an open question raised by Yuan et al. \cite{Yuan}, which fills a gap in the study of strongly nonlocal OPS.

Recently,  a weaker form of nonlocality which is called local distinguishability based genuine nonlocality were considered in \cite{Rout21,Li21}. A multipartite OPS 
 is said to be genuinely nonlocal if it is
locally indistinguishable for every bipartition of the subsystems. 
The Propositions 1-2 of Ref. \cite{Rout21} and the Propositions 1-3 of Ref. \cite{Li21} showed several genuinely nonlocal OPSs in multipartite systems. 
Noting that a locally irreducible set is locally indistinguishable, but a locally indistinguishable set in general is not locally irreducible except when it contains three orthogonal pure states \cite{Halder}.
Since a strongly nonlocal OPS is an orthogonal set of fully product
states in multipartite quantum systems that is locally irreducible for every
bipartition of its subsystems, a strongly nonlocal OPS is clearly genuinely nonlocal, but the converse does not always hold.
The strongly nonlocal OPSs in Theorems 2-3 are genuinely nonlocal, while the genuinely nonlocal OPSs in Propositions 1-2 of Ref. \cite{Rout21} and the Propositions 1-3 of Ref. \cite{Li21} are not strongly nonlocal.

\section{entanglement-assisted discrimination}\label{Q3}
It is impossible that a strongly nonlocal orthogonal set is perfected discriminated by LOCC even if any $n-1$ parties are allowed together unless there is enough entanglement resource. In this section, according to the method in Ref. \cite{ZhangWZ,Rout,ZGY}, we consider the entanglement-assisted discrimination of our OPSs when all parties are separated.

\emph{Theorem 4}. In $\otimes_{i=1}^{n}\mathcal{C}^{3}$ $(n>3)$, the set $\mathcal{E}$ can be perfectly distinguished by LOCC with $n-1$ copies of $3\otimes 3$ maximally entangled state, where $n$ is even.

The discrimination protocol proceeds as follows. Let the first subsystem be denoted as Alice, and the remaining subsystems $2,\ldots,n$ are denoted as $\textup{Bob}1,\ldots,\textup{Bob}(n-1)$, respectively. $|\phi^{+}(3)\rangle=\frac{1}{\sqrt{3}}\sum_{i=0}^{2}|ii\rangle$ is a maximally entangled state in $\mathcal{C}^{3}\otimes \mathcal{C}^{3}$. Alice shares the maximally entangled states $|\phi^{+}(3)\rangle_{a_{1}b_{1}},\ldots,|\phi^{+}(3)\rangle_{a_{(n-1)}b_{(n-1)}}$ with $\textup{Bob}1,\ldots,\textup{Bob}(n-1)$, respectively. Then, the initial subset is
\begin{equation}
\mathcal{P}_{Q}\otimes|\phi^{+}(3)\rangle_{a_{1}b_{1}}\otimes\cdots\otimes|\phi^{+}(3)\rangle_{a_{(n-1)}b_{(n-1)}},
\end{equation}
where $\mathcal{P}_{Q}$ $(Q\in \Theta,~\mathcal{P}\in \{\mathcal{C},\mathcal{D}\})$ is a subset of $\mathcal{E}$ and $a_{1}\cdots a_{(n-1)},~b_{1},\ldots,b_{(n-1)}$ are ancillary systems of Alice, $\textup{Bob}1,\ldots,\textup{Bob}(n-1)$, respectively. Because the states in each subset $\mathcal{P}_{Q}$ is locally distinguishable, we only need to identify these subsets by LOCC. To do this, we take the following two steps.

$Step~1.$ Each of $\textup{Bob}j$ $(j=1,\ldots,n-1)$ performs the corresponding measurement
\begin{equation}
\begin{aligned}
\mathcal{M}_{j}\equiv\{&M_{j1}:=P[|0\rangle_{B_{j}};|0\rangle_{b_{j}}]+P[|1\rangle_{B_{j}};|1\rangle_{b_{j}}]\\
                       &\quad\quad\quad~+P[|2\rangle_{B_{j}};|2\rangle_{b_{j}}],\\
                       &M_{j2}:=P[|0\rangle_{B_{j}};|1\rangle_{b_{j}}]+P[|1\rangle_{B_{j}};|2\rangle_{b_{j}}]\\
                       &\quad\quad\quad~+P[|2\rangle_{B_{j}};|0\rangle_{b_{j}}],\\
                       &M_{j3}:=I-M_{j1}-M_{j2}\}.
\end{aligned}
\end{equation}
Here and below, $P[(|k\rangle,\ldots,|l\rangle)_{X};|m\rangle_{Y}]:=(|k\rangle\langle k|+\cdots+|l\rangle\langle l|)_{X}\otimes|m\rangle\langle m|_{Y}$ for the states $|k\rangle,\ldots,|m\rangle$ and the corresponding subsystems $X$ and $Y$. Suppose $M_{11},~M_{21},\ldots,M_{(n-1)1}$ click, the resulting postmeasurement subset is
\begin{equation}
\widetilde{\mathcal{P}_{Q}}=\mathcal{P}_{Q}^{(1)}\otimes\mathcal{P}_{Q}^{(\tilde{2})}\otimes\cdots\otimes\mathcal{P}_{Q}^{(\tilde{n})},
\end{equation}
where $\mathcal{P}_{Q}^{(\widetilde{j+1})}$ $(j=1,\ldots,n-1)$ is equal to the following
\begin{equation}
\left\{
\begin{aligned}
&|0\rangle_{B_{j}}|00\rangle_{a_{j}b_{j}},&& \textup{if}~\mathcal{P}_{Q}^{(j+1)}=|0\rangle_{B_{j}},\\
&|0\rangle_{B_{j}}|00\rangle_{a_{j}b_{j}}\pm |1\rangle_{B_{j}}|11\rangle_{a_{j}b_{j}},&&\textup{if}~\mathcal{P}_{Q}^{(j+1)}=|\eta_{\pm}\rangle_{B_{j}},\\
&|1\rangle_{B_{j}}|11\rangle_{a_{j}b_{j}}\pm |2\rangle_{B_{j}}|22\rangle_{a_{j}b_{j}},&&\textup{if}~\mathcal{P}_{Q}^{(j+1)}=|\xi_{\pm}\rangle_{B_{j}},\\
&|2\rangle_{B_{j}}|22\rangle_{a_{j}b_{j}},&&\textup{if}~\mathcal{P}_{Q}^{(j+1)}=|2\rangle_{B_{j}}.
\end{aligned}
\right.
\end{equation}
Then, we go to the next step.

$Step~2.$ Alice performs the measurement
\begin{equation}
\begin{aligned}
\mathcal{M}_{n}\equiv\{&M_{\mathcal{P},Q}:=P[\mathcal{P}_{Q}^{[1]};\mathcal{P}_{Q}^{[2]}|_{a_{1}};\ldots;\mathcal{P}_{Q}^{[n]}|_{a_{n-1}}],\\
&\mathcal{M}_{n1}:=P[|1\rangle_{A};|1\rangle_{a_{1}};\ldots;|1\rangle_{a_{n-1}}]\}_{\mathcal{P},Q}.
\end{aligned}
\end{equation}
Here $\mathcal{P}_{Q}^{[1]}$ (\ref{14}) is a subset of computational basis on the subsystem $A$ and
\begin{equation}
\mathcal{P}_{Q}^{[j+1]}|_{a_{j}}=
\left\{
\begin{aligned}
&|0\rangle_{a_{j}},&& \textup{if}~\mathcal{P}_{Q}^{(j+1)}=|0\rangle_{B_{j}},\\
&(|0\rangle,|1\rangle)_{a_{j}},&&\textup{if}~\mathcal{P}_{Q}^{(j+1)}=|\eta_{\pm}\rangle_{B_{j}},\\
&(|1\rangle,|2\rangle)_{a_{j}},&&\textup{if}~\mathcal{P}_{Q}^{(j+1)}=|\xi_{\pm}\rangle_{B_{j}},\\
&|2\rangle_{a_{j}},&&\textup{if}~\mathcal{P}_{Q}^{(j+1)}=|2\rangle_{B_{j}},
\end{aligned}
\right.
\end{equation}
for $j=1,\ldots,n-1$. It is obvious that the projective measurement $\mathcal{M}_{n}$ satisfies the completeness equation $\sum_{\mathcal{P},Q}M_{\mathcal{P},Q}^{\dagger}M_{\mathcal{P},Q}+\mathcal{M}_{n1}^{\dagger}\mathcal{M}_{n1}=I$ in Ref. \cite{Nielsen}. Meanwhile, the resulting postmeasurement subset is $\widetilde{\mathcal{P}_{Q}}$ with respect to every $M_{\mathcal{P},Q}$. That is, we have $M_{\mathcal{P},Q}\widetilde{\mathcal{P}_{Q}}=\widetilde{\mathcal{P}_{Q}}$, while $M_{\mathcal{P},Q}\widetilde{\mathcal{P}'_{Q'}}=0$ when $\mathcal{P}'\neq \mathcal{P}$ or $Q'\neq Q$. Here $Q'\in\Theta$ and $\mathcal{P}'\in \{\mathcal{C},\mathcal{D}\}$.

If other POVM elements click in step 1, we can find similar protocol to perfectly LOCC distinguish these states.

Because of the same structure as $\mathcal{E}$, we give the conclusion for OPS $\mathcal{F}$.

\emph{Theorem 5}. In $\otimes_{i=1}^{n}\mathcal{C}^{d_{i}}$ $(n>3,d_{i}\geq 3)$, $n-1$ copies of $3\otimes 3$ maximally entangled state are enough to perfectly LOCC distinguish set $\mathcal{F}$, where $n$ is even.

The local discrimination protocol for set $\mathcal{F}$ is similar to that above. We just need to replace $|1\rangle_{i}$ and $|2\rangle_{i}$ $(i=1,\ldots,n)$ in the measurements with $(|1\rangle,|2\rangle,\ldots,|d_{i}-2\rangle)_{i}$ and $|d_{i}-1\rangle_{i}$, respectively. Here $d_{i}$ is the dimension of the $i$th subsystem. That is, the measurements in steps 1 and 2 become
\begin{equation}
\begin{aligned}
\mathcal{M}_{j}\equiv\{&M_{j1}:=P[|0\rangle_{B_{j}};|0\rangle_{b_{j}}]+P[|d_{j+1}-1\rangle_{B_{j}};|2\rangle_{b_{j}}]\\
                       &\quad\quad\quad~+P[(|1\rangle,\ldots,|d_{j+1}-2\rangle)_{B_{j}};|1\rangle_{b_{j}}],\\
                       &M_{j2}:=P[|0\rangle_{B_{j}};|1\rangle_{b_{j}}]+P[|d_{j+1}-1\rangle_{B_{j}};|0\rangle_{b_{j}}]\\
                       &\quad\quad\quad~+P[(|1\rangle,\ldots,|d_{j+1}-2\rangle)_{B_{j}};|2\rangle_{b_{j}}],\\
                       &M_{j3}:=I-M_{j1}-M_{j2}\},
\end{aligned}
\end{equation}
for $j=1,\ldots,n-1$ and
\begin{equation}
\begin{aligned}
\mathcal{M}_{n}\equiv\{&M_{\mathcal{P},Q}:=P[\mathcal{P}_{Q}^{f[1]};\mathcal{P}_{Q}^{[2]}|_{a_{1}};\ldots;\mathcal{P}_{Q}^{[n]}|_{a_{n-1}}],\\
&\mathcal{M}_{n1}:=P[(|1\rangle,|2\rangle,\ldots,|d_{1}-2\rangle)_{A};|1\rangle_{a_{1}};\\
&\quad\quad\quad~~~\ldots;|1\rangle_{a_{n-1}}\}_{\mathcal{P},Q},
\end{aligned}
\end{equation}
respectively. Meanwhile, $\mathcal{P}_{Q}^{f[1]}$ is a projection set of $\mathcal{P}_{Q}^{f(1)}$ on party $A$, i.e.,
\begin{equation}
\mathcal{P}_{Q}^{f[1]}=
\left\{
\begin{aligned}
&|0\rangle_{A},&& \textup{if}~\mathcal{P}_{Q}^{f(1)}=|0\rangle_{A},\\
&(|0\rangle,\ldots,|d_{1}-2\rangle)_{A},&&\textup{if}~\mathcal{P}_{Q}^{f(1)}=\{|\alpha_{k}\rangle_{A}\},\\
&(|1\rangle,\ldots,|d_{1}-1\rangle)_{A},&&\textup{if}~\mathcal{P}_{Q}^{f(1)}=\{|\beta_{l}\rangle_{A}\},\\
&|d_{1}-1\rangle_{A},&&\textup{if}~\mathcal{P}_{Q}^{f(1)}=|d_{1}-1\rangle_{A}.
\end{aligned}
\right.
\end{equation}

In any finite dimensional system, our scheme consumes $\log_{2}3^{n-1}$ ebits entanglement resource in total, while the quantum teleportation scheme needs $\log_{2}d_{2}\cdots d_{n}$ ebits. Here we suppose $d_{1}=\max\{d_{1},\ldots,d_{n}\}$ without loss of generality. Compared to the teleportation, our method consumes the same entanglement resource in the system $\mathcal{C}^{d_{1}}\otimes(\otimes_{i=2}^{n}\mathcal{C}^{3})$ $(n>3)$, but it consumes less resource in the other high-dimensional systems.

\section{Information hiding in multipartite quantum states}\label{Q4}
Local indistinguishability can be used for information hiding. In this section, we discuss the value of strong nonlocal OPSs for quantum data hiding in multipartite quantum states \cite{DiVincenzo,Eggeling,ShiYCZ}. In a situation, a boss has a set of data on which she would like $n$ subordinates to operate by LOCC without letting them know the sensitive data \cite{DiVincenzo,ShiYCZ}. Assume that these subordinates are far away from each other and can only perform LOCC. To solve this problem, the boss can encode this piece of data in an $n$-partite OPS with strong quantum nonlocality and then send it to his subordinates. Due to the irreducibility of this OPS in every bipartition, these subordinates cannot access to information at all even if $n-1$ of them are collusive. When $n$ is even, the information can be encoded in our set $\mathcal{F}$, while when $n$ is odd, information can be encoded in set $\mathcal{O}$ of Ref. \cite{He}. Compared to using orthogonal entangled sets with strong quantum nonlocality \cite{ShiYCZ}, one obvious advantage of strongly nonlocal OPSs is that they do not need any entangled resources. Still considering this scenario, we add an auxiliary system. That is, the boss can encode the information in $(n+1)$-partite OPS with strong quantum nonlocality and the former $n$ subsystems are programmed with real information while the $(n+1)$th subsystem does not put any meaningful information. Then, the boss send the former $n$ subsystems to his subordinates and master the $(n+1)$th subsystem. No matter what measurements are performed on the former $n$ subsystems, the $(n+1)$th subsystem will not be affected. Moreover, the information will not be accessible at all even if all these subordinates conspire.

\section{Conclusion}\label{Q5}
In summary, we construct the OPSs with strong quantum nonlocality in any possible $n$-partite systems, where $n$ is even, based on the structure introduced by He et al. \cite{He}. We analyze the differences and connections between the sets given by He et al. \cite{He} and our sets. Our results together with  Ref. \cite{He}  jointly answer the open questions in Ref. \cite{Halder} and Ref. \cite{Yuan}, in a word, there is a general construction of incomplete orthogonal product bases with strong nonlocality in any possible systems. Not only that, we also provide an entanglement-assisted local discrimination protocol for our constructed sets, which is resource efficient as it consumes less entanglement in comparison with the teleportaion-based protocol when at least two subsystems have dimensions greater than three. Finally, we connected strongly nonlocal OPSs with quantum data hiding in multipartite quantum states.

\begin{acknowledgments}
This work was supported by the National Natural Science Foundation of China under Grant Nos. 12071110 and 62271189, the Hebei Natural Science Foundation of China under Grant No. A2020205014, and funded by the Science and Technology Project of Hebei Education Department under Grant Nos. ZD2020167 and ZD2021066.
\end{acknowledgments}

\begin{appendix}
\section{Remove the $(n+1)$th subsystem of set $\mathcal{O}$}\label{A}
In $(n+1)$-qutrit quantum system, let us first consider the case where every index $K$ contains $n$ elements. Here $n$ is even. Obviously, there are $C_{n+1}^{n}$ such indexes $K$ and $C_{n}^{n}$ index $K$ without element $n+1$. We get rid of the index $K$ without element $n+1$. Then we remove the element $n+1$ in the remaining $K$. At this point, each of the resulting new indexes $K'$ is an arbitrary combination of $n-1$ elements in $I_{n}$ and there are $C_{n}^{n-1}=C_{n+1}^{n}-C_{n}^{n}$ indexes $K'$. This happens to be all the indexes $Q$ which contain $n-1$ elements in $n$-qutrit quantum system. Similarly, we have same conclusion for the $K$ containing $2,4,\ldots,n-2$ elements, respectively. So, the new set $\mathcal{E}_{2}$ is the same as the set $\mathcal{E}$. By the same way, we can know that the set $\mathcal{O}$ in $(n-1)$-qutrit quantum system can be obtained by removing the $n$th subsystem of set $\mathcal{E}$.

\section{The relationship between the sets $\mathcal{E}$ and $\mathcal{O}$}\label{B}
The cyclic permutation of the parties is defined as
\begin{equation*}
\begin{aligned}
 &P_{j}^{C}(|\psi\rangle_{1}|\psi\rangle_{2}\cdots|\psi\rangle_{n+1})\\
=&|\psi\rangle_{j}|\psi\rangle_{j+1}\cdots|\psi\rangle_{n+1}|\psi\rangle_{1}|\psi\rangle_{2}\cdots|\psi\rangle_{j-1},
\end{aligned}
\end{equation*}
for $j\in \mathcal{Z}_{n+1}$. Here $n$ is even. Evidently $P_{1}^{C}$ is an identity permutation. Let $(k_{1}k_{2}\cdots k_{n+1})$ express an arrangement of parties, where $k_{i}\in I_{n+1}$.

We consider the set $\mathcal{O}$ on arrangement $(12\cdots(n+1))$. Because of the symmetry, for a fixed $i\in I_{n}$, set $P_{i+1}^{C}(\mathcal{O})$ is the same as set $\mathcal{O}$. Meanwhile, it corresponds to the arrangement $((i+1)\cdots(n+1)1\cdots i)$. When we remove the $i$th subsystem of set $P_{i+1}^{C}(\mathcal{O})$, the new set is $\mathcal{E}$ on the arrangement $((i+1)\cdots(n+1)1\cdots (i-1))$, according to the known result above. Hence, the set $\mathcal{O}$ removing the $i$th subsystem is the set $P_{n+2-i}^{C}(\mathcal{E})$ on the arrangement $(1\cdots(i-1)(i+1)\cdots (n+1))$.

Given the set $\mathcal{E}$ on arrangement $(12\cdots n)$, we consider to remove its $i$th subsystem where $i\in I_{n-1}$. We first consider the $Q$ containing $n-1$ elements. There are $C_{n}^{n-1}$ such indexes $Q$. We remove the index $Q$ without element $i$ and there are $C_{n-1}^{n-1}$ such indexes in total. Then, we remove the element $i$ in the remaining $Q$. Each new index $Q'$ is a combination of $n-2$ elements in $\{1,\ldots, i-1,i+1,\ldots,n\}$ and has $C_{n-1}^{n-2}=C_{n}^{n-1}-C_{n-1}^{n-1}$ such indexes. We have similar conclusion for the $Q$ containing $1,3,\ldots,n-3$ elements, respectively. They can exactly correspond to all indexes $K$ belonging to $\Lambda$ in $(n-1)$-qutrit system. Suppose that $\mathcal{O}_{2}$ expresses the new set after we remove the $i$th subsystem of set $\mathcal{E}$. According to the construction method of sets $\mathcal{E}$ and $\mathcal{O}$, to turn $\mathcal{O}_{2}$ to $\mathcal{O}$, we only need to transform $|0\rangle_{j}$, $|\eta_{\pm}\rangle_{j}$, $|\xi_{\pm}\rangle_{j}$ and $|2\rangle_{j}$ to $|2\rangle_{j}$, $|\xi_{\pm}\rangle_{j}$, $|\eta_{\pm}\rangle_{j}$ and $|0\rangle_{j}$ for all $j>i$, respectively.

\section{The proof of theorem 2}\label{D}
Before proving the theorem, we need to introduce some notations and concepts given in \cite{ZGY}. Let $X=\{l_{1},\ldots,l_{m}\}$ denote a subset of $I_{n}$. Assume $\mathcal{P}_{Q}^{[X]}=\mathcal{P}_{Q}^{[l_{1}]}\otimes\cdots\otimes\mathcal{P}_{Q}^{[l_{m}]}$ expresses the projection set of subset $\mathcal{P}_{Q}$ on party $X$. About the subset $\mathcal{P}_{Q}$, if there are some subsets $\mathcal{P}_{Q}'$ being different from it and satisfying $\mathcal{P}_{Q}^{[X]}\subset \cup \mathcal{P}_{Q}'^{[X]}$ and $\cap \mathcal{P}_{Q}'^{[\bar{X}]}\neq \emptyset$, where $\bar{X}$ is the complement of $X$, then the union $R_{\mathcal{P},Q}=\cup\mathcal{P}_{Q}'$ is called the projection inclusion (PI) set of $\mathcal{P}_{Q}$ on party $X$. Specially, if there exists a subset $\mathcal{P}_{Q}'\subset R_{\mathcal{P},Q}$ such that $|\mathcal{P}_{Q}^{[X]}\cap \mathcal{P}_{Q}'^{[X]}|=1$, then $R_{\mathcal{P},Q}$ is called a more useful projection inclusion (UPI) set \cite{ZGY}.

We introduce a set sequence $G_{1},G_{2},\ldots,G_{s}$ about set $\mathcal{E}$ \cite{ZGY}. It is a partition of set $\mathcal{E}$. Here each set $G_{x}$ $(x=1,\ldots,s)$ is a union of some different subsets $\mathcal{P}_{Q}$ and they satisfy the following three conditions.

1) The set $G_{1}$ is the union of all subsets that have UPI sets.

2) The intersection of any two sets in this set sequence is empty set and the union $\cup_{x=1}^{s}G_{x}$ is equal to the set $\mathcal{E}$.

3) For any $\mathcal{P}_{Q_{x+1}}$ contained in $G_{x+1}$, there is at least one $\mathcal{P}_{Q_{x}}$ contained in $G_{x}$ such that $\mathcal{P}_{Q_{x+1}}^{[X]}\cap \mathcal{P}_{Q_{x}}^{[X]}\neq\emptyset$.

Let $\mathcal{B}^{X}=\{|i\rangle_{X}\}_{i=0}^{d_{X}-1}=\{\otimes_{k=1}^{m} |i_{l_{k}}\rangle|i_{l_{k}}=0,\ldots,d_{l_{k}}-1\}$ represent the computational basis of $X$ party, where $d_{X}=d_{l_{1}}\cdots d_{l_{m}}$.
For each $i\in\mathcal{Z}_{d_{X}}$, we define $\mathcal{B}_{i}^{X}:=\{|j\rangle_{X}\}_{j=i}^{d_{X}-1}$, $V_{i}:=\{\cup\mathcal{P}_{Q}^{[\bar{X}]}~|~|i\rangle_{X}\in \mathcal{P}_{Q}^{[X]}\}$ and $\widetilde{S}_{V_{i}}:=\{\cup\mathcal{P}_{Q}^{[X]}~|~\mathcal{P}_{Q}^{[\bar{X}]}\cap V_{i}\neq \emptyset\}$. Ref. \cite{ZGY} has the detailed examples to explain these symbols. Next we will prove Theorem 2 by using \cite[Theorem 1]{ZGY}.

We first consider the orthogonality-preserving POVM performed on party $X_{1}$ $(=I_{n}\setminus \{1\})$. To prove that it can only be trivial, we only need to show that the four conditions in \cite[Theorem 1]{ZGY} are satisfied. For simplicity, let $X$ express $X_{1}$. The specific conditions are described as

i) For any $i\in\mathcal{Z}_{d_{X}-1}$, there is a relationship $\mathcal{B}_{i}^{X}\subset \widetilde{S}_{V_{i}}$,

ii) For any subset $\mathcal{P}_{Q}$, there exists a corresponding PI set $R_{\mathcal{P},Q}$ on party $X$,

iii) There is a set sequence $G_{1},\ldots,G_{s}$ satisfying 1)-3). For each $\mathcal{P}_{Q_{x+1}}\subset G_{x+1}$ with $x=1,\ldots,s-1$, there exist a $\mathcal{P}_{Q_{x}}\subset G_{x}$ and a $\mathcal{P}_{Q}'\subset R_{\mathcal{P},Q}$ such that $\mathcal{P}_{Q_{x+1}}^{[X]}\cap \mathcal{P}_{Q_{x}}^{[X]}\supset \mathcal{P}_{Q_{x+1}}^{[X]}\cap \mathcal{P}_{Q}'^{[X]}$,

iv) The family of sets $\{\mathcal{P}_{Q}^{[X]}\}_{\mathcal{P}_{Q}\subset \mathcal{E}}$ is connected.

In $1|X$ bipartition, the plane structure of the set $\mathcal{E}$ is shown in Fig. \ref{11}. By observing this tile graph, the conditions i) and ii) are obvious.

When $Q=\{1\}$, we have $\mathcal{C}_{Q}=\{|\eta_{\pm}\rangle_{1}|0\rangle_{2}\cdots|0\rangle_{n}\}$. So, there is at least one subset which has UPI set. If condition iv) holds, we have the set sequence $\{G_{1},G_{2},\ldots,G_{s}\}$. Because of the special plane structure, the set $R_{D}=\cup_{Q\in \Theta}\mathcal{D}_{Q}$ is a PI set of any subset $\mathcal{C}_{Q}$. Corresponding, the set $R_{C}=\cup_{Q\in \Theta}\mathcal{C}_{Q}$ is a PI set of any subset $\mathcal{D}_{Q}$. For any subset $\mathcal{C}_{Q_{x+1}}$ $(\textup{or}~\mathcal{D}_{Q_{x+1}})$ of $G_{x+1}$ $(x=1,\ldots,s-1)$, there exists a corresponding subset $\mathcal{D}_{Q_{x}}$ $(\textup{or}~\mathcal{C}_{Q_{x}})$ of set $G_{x}$ such that $\mathcal{D}_{Q_{x}}^{[X]}\cap \mathcal{C}_{Q_{x+1}}^{[X]} \neq \emptyset$ $(\textup{or}~\mathcal{C}_{Q_{x}}^{[X]}\cap \mathcal{D}_{Q_{x+1}}^{[X]} \neq \emptyset)$. Meanwhile, the subset $\mathcal{D}_{Q_{x}}$ $(\textup{or}~\mathcal{C}_{Q_{x}})$ is also contained in the set $R_{D}$ $(\textup{or}~R_{C})$. This means $G_{x}\cap R_{D}\neq \emptyset$ $(\textup{or}~G_{x}\cap R_{C}\neq \emptyset)$. Obviously, these subsets satisfy the relationship of condition iii). That is, the condition iii) holds when condition iv) is satisfied.

About the set $\mathcal{O}$ in $\otimes_{i=2}^{n}\mathcal{C}^{3}$ system, He et al. \cite{He} have proven that any orthogonality-preserving POVM performed on $X_{12}$ $(=\{3,\ldots,n\})$ party can only be trivial. According to the \cite[Corollary 2]{ZGY}, the set $\cup_{\mathcal{P}_{K}\subset \mathcal{O}}\mathcal{P}_{K}^{[X_{12}]}$ is the computational basis $\mathcal{B}^{X_{12}}$ corresponding to subsystem $X_{12}$ and the family of projection sets $\{\mathcal{P}_{K}^{[X_{12}]}\}_{\mathcal{P}_{K}\subset \mathcal{O}}$ is connected. Here the index set $K$ expresses a subset of $I_{n}\setminus \{1\}$.

\begin{figure}[h]
\centering
\includegraphics[width=0.4\textwidth]{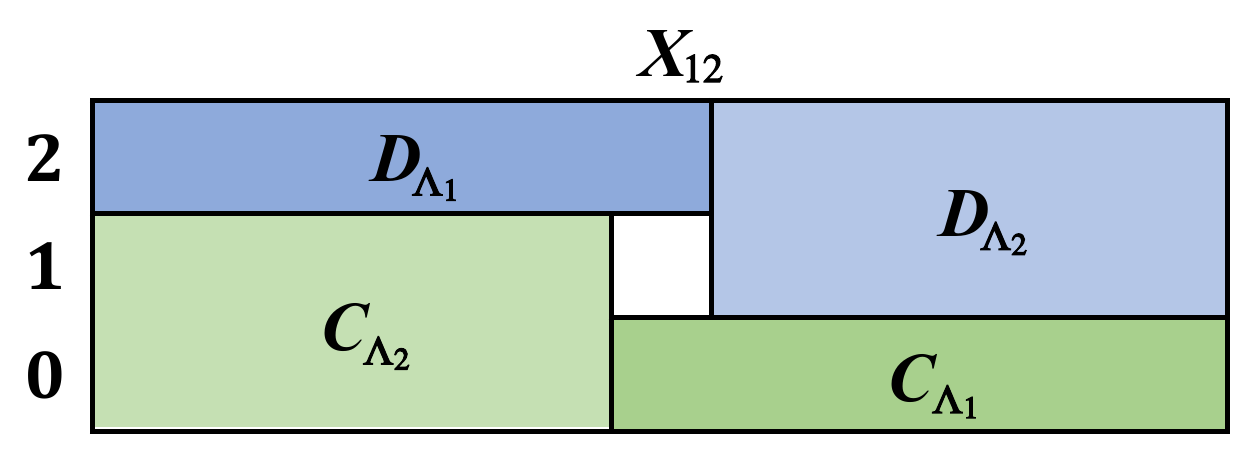}
\caption{This is the plane tile of set $\mathcal{O}$ in $2|X_{12}$ bipartition. Here, the projection sets of $\mathcal{C}_{\Lambda_{1}}$ (or $\mathcal{C}_{\Lambda_{2}}$) on the $\{2\}$ party and $X_{12}$ party are $\{|0\rangle_{2}\}$ (or $\{|0\rangle_{2},|1\rangle_{2}\}$) and $\bigcup_{K\in\Lambda_{1}}\mathcal{C}_{K}^{[X_{12}]}$ (or $\bigcup_{K\in\Lambda_{2}}\mathcal{C}_{K}^{[X_{12}]}$), respectively. Moreover, the intersection of $\bigcup_{K\in\Lambda_{1}}\mathcal{C}_{K}^{[X_{12}]}$ and $\bigcup_{K\in\Lambda_{2}}\mathcal{C}_{K}^{[X_{12}]}$ is the empty set and the union of them is the computational basis on party $X_{12}$. It is similar for $\mathcal{D}_{\Lambda_{1}}$ and $\mathcal{D}_{\Lambda_{2}}$. \label{12}}
\end{figure}

For convenience, we use $\mathcal{P}_{K}^{(X_{12})}$ to denote $\mathcal{P}_{K}^{(3)}\otimes\cdots\otimes\mathcal{P}_{K}^{(n)}$. Let $\mathcal{C}_{\Lambda_{1}}=|0\rangle_{2}\otimes\mathcal{C}_{\Lambda_{1}}^{(X_{12})}=\{|0\rangle_{2}\otimes\mathcal{C}_{K}^{(X_{12})}\}_{K\in \Lambda_{1}}$ and $\mathcal{C}_{\Lambda_{2}}=|\eta_{\pm}\rangle_{2}\otimes\mathcal{C}_{\Lambda_{2}}^{(X_{12})}=\{|\eta_{\pm}\rangle_{2}\otimes\mathcal{C}_{K}^{(X_{12})}\}_{K\in \Lambda_{2}}$, where $\Lambda_{1}=\{K|K\in\Lambda~\textup{and}~2\notin K\}$ and $\Lambda_{2}=\Lambda\setminus \Lambda_{1}$. Correspondingly, we have $\mathcal{D}_{\Lambda_{1}}=|2\rangle_{2}\otimes\mathcal{D}_{\Lambda_{1}}^{(X_{12})}$ and $\mathcal{D}_{\Lambda_{2}}=|\xi_{\pm}\rangle_{2}\otimes\mathcal{D}_{\Lambda_{2}}^{(X_{12})}$. From this, we give the plane tile of set $\mathcal{O}$ in $2|X_{12}$ bipartition as shown in Fig. \ref{12}.

\begin{figure}[h]
\centering
\includegraphics[width=0.48\textwidth]{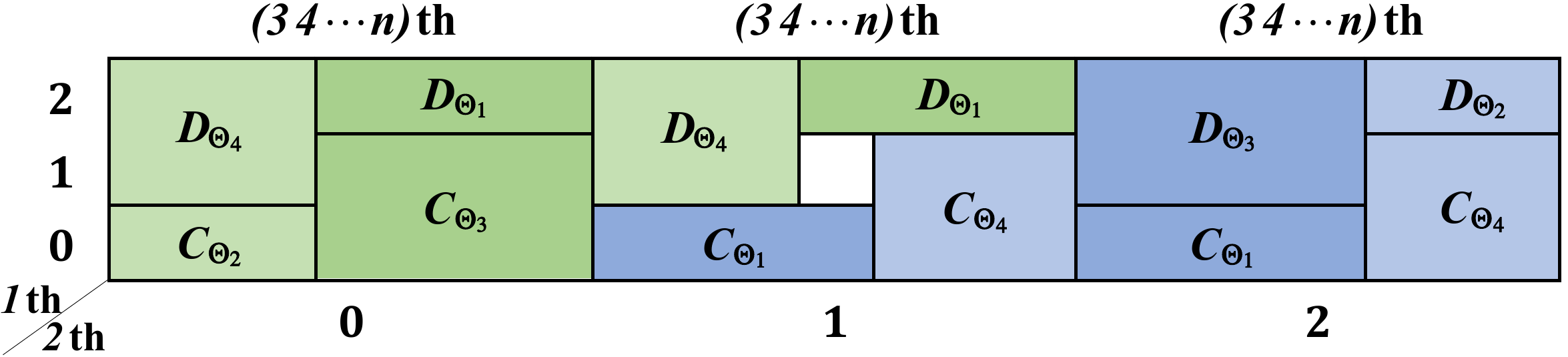}
\caption{This is the plane structure of the set $\mathcal{E}$ in $1|X$ bipartition. Here $i$th $(i=1,\ldots,n)$ expresses the $i$th subsystem and $(34\cdots n)$th means the $X_{12}$ party. The projection sets of $\mathcal{D}_{\Theta_{1}}$ on the $\{1\}$ party, $\{2\}$ party and $X_{12}$ party are $\{|2\rangle_{1}\}$, $\{|0\rangle_{2},|1\rangle_{2}\}$ and $\cup_{K\in\Lambda_{1}}\mathcal{C}_{K}^{[X_{12}]}$, respectively. In particular, the projection set of $\mathcal{D}_{\Theta_{1}}$ on the $X_{12}$ party is the same as the projection set of $\mathcal{C}_{\Lambda_{1}}$ in Fig. \ref{12} on the $X_{12}$ party. The other sets in this diagram have similar consideration. \label{13}}
\end{figure}

According to the relationship between the sets $\mathcal{E}$ and $\mathcal{O}$, it is easy to know that
\begin{equation*}
\begin{aligned}
&\mathcal{C}_{\Theta_{1}}=|0\rangle\otimes|\xi_{\pm}\rangle\otimes\mathcal{D}_{\Lambda_{1}}^{(X_{12})},\\
&\mathcal{C}_{\Theta_{2}}=|0\rangle\otimes|0\rangle\otimes\mathcal{C}_{\Lambda_{2}}^{(X_{12})},\\
&\mathcal{C}_{\Theta_{3}}=|\eta_{\pm}\rangle\otimes|0\rangle\otimes\mathcal{C}_{\Lambda_{1}}^{(X_{12})},\\
&\mathcal{C}_{\Theta_{4}}=|\eta_{\pm}\rangle\otimes|\xi_{\pm}\rangle\otimes\mathcal{D}_{\Lambda_{2}}^{(X_{12})},\\
&\mathcal{D}_{\Theta_{1}}=|2\rangle\otimes|\eta_{\pm}\rangle\otimes\mathcal{C}_{\Lambda_{1}}^{(X_{12})}, \\
&\mathcal{D}_{\Theta_{2}}=|2\rangle\otimes|2\rangle\otimes\mathcal{D}_{\Lambda_{2}}^{(X_{12})}, \\
&\mathcal{D}_{\Theta_{3}}=|\xi_{\pm}\rangle\otimes|2\rangle\otimes\mathcal{D}_{\Lambda_{1}}^{(X_{12})}, \\
&\mathcal{D}_{\Theta_{4}}=|\xi_{\pm}\rangle\otimes|\eta_{\pm}\rangle\otimes\mathcal{C}_{\Lambda_{2}}^{(X_{12})}.
\end{aligned}
\end{equation*}
Here $\Theta_{1}=\{Q\in\Theta|1\notin Q~\textup{and}~2\in Q\}$, $\Theta_{2}=\{Q\in\Theta|1,2\notin Q\}$, $\Theta_{3}=\{Q\in\Theta|1\in Q~\textup{and}~2\notin Q\}$, $\Theta_{4}=\{Q\in\Theta|1,2\in Q\}$. Then the plane structure of the set $\mathcal{E}$ also can be shown as Fig. \ref{13}.

Due to the connectedness of the family of projection sets $\{\mathcal{P}_{K}^{[X_{12}]}\}_{\mathcal{P}_{K}\subset \mathcal{O}}$, it is not difficult to know that the group of projection sets $\{\mathcal{D}_{\Theta_{1}}^{[X]},\mathcal{C}_{\Theta_{1}}^{[X]},\mathcal{D}_{\Theta_{4}}^{[X]},\mathcal{C}_{\Theta_{4}}^{[X]}\}$ is connected. Meanwhile, we find that the union of these four projection sets is the computational basis on party $X$. That is, the family of projection sets $\{\mathcal{P}_{Q}^{[X]}\}_{\mathcal{P}_{Q}\subset \mathcal{E}}$ cannot be divided into two groups of sets $\{\mathcal{P}_{Q}^{[X]}\}_{\mathcal{P}_{Q}\subset \mathcal{E}'}$ $(\mathcal{E}'\subsetneqq \mathcal{E})$ and $\{\mathcal{P}_{Q}^{[X]}\}_{\mathcal{P}_{Q}\subset \mathcal{E}\setminus\mathcal{E}'}$ such that
\begin{equation*}
\begin{aligned}
\left(\bigcup\limits_{\mathcal{P}_{Q}\subset \mathcal{E}'}\mathcal{P}_{Q}^{[X]}\right)\bigcap \left(\bigcup\limits_{\mathcal{P}_{Q}\subset \mathcal{E}\setminus\mathcal{E}'}\mathcal{P}_{Q}^{[X]}\right)=\emptyset.
\end{aligned}
\end{equation*}
So, the family of projection sets $\{\mathcal{P}_{Q}^{[X]}\}_{\mathcal{P}_{Q}\subset \mathcal{E}}$ is connected. the condition iv) is shown.

Therefore, any orthogonality-preserving POVM element performed on party $X$ can only be proportional to the identity operator. For the other subsystems $X_{2},\ldots,X_{n}$, by the same way, it is easy to illustrate that any orthogonality-preserving POVM performed on party $X_{i}$ $(i=2,\ldots,n)$ can only be trivial. The proof of this theorem is completed.

\end{appendix}


\begin{thebibliography}{99}

%1-10
\bibitem{Einstein} A. Einstein, B. Podolsky, and N. Rosen, Can quantum-mechanical description of physical reality be considered complete?, \href{https://journals.aps.org/pr/abstract/10.1103/PhysRev.47.777} {Phys. Rev. \textbf{47}, 777 (1935)}.

\bibitem{Werner} R. F. Werner, Quantum states with Einstein-Podolsky-Rosen correlations admitting a hidden-variable model, \href{https://journals.aps.org/pra/abstract/10.1103/PhysRevA.40.4277} {Phys. Rev. A \textbf{40}, 4277 (1989)}.

\bibitem{Horodecki} R. Horodecki, P. Horodecki, M. Horodecki, and K. Horodecki, Quantum entanglement, \href{https://doi.org/10.1103/RevModPhys.81.865} {Rev. Mod. Phys. \textbf{81}, 865 (2009)}.

\bibitem{GYvE} T. Gao, F. L. Yan, and S. J. van Enk, Permutationally invariant part of a density matrix and nonseparability of $N$-qubit states, \href{https://journals.aps.org/prl/abstract/10.1103/PhysRevLett.112.180501} {Phys. Rev. Lett. \textbf{112}, 180501 (2014)}.

\bibitem{BennettFS} C. H. Bennett, C. A. Fuchs, and J. A. Smolin, \textit{Quantum Communication, Computing, and Measurement} (Plenum, New York, 1997).

\bibitem{Buhrman} H. Buhrman, R. Cleve, and A. Wigderson, \textit{Proceedings of the 30th Annual ACM Symposium on the Theory of Computing} (ACM, Los Alamitos, 1998).

\bibitem{Deng} F. G. Deng, G. L. Long, and X. S. Liu, Two-step quantum direct communication protocol using the Einstein-Podolsky-Rosen pair block, \href{https://journals.aps.org/pra/abstract/10.1103/PhysRevA.68.042317} {Phys. Rev. A \textbf{68}, 042317 (2003)}.

\bibitem{GaoYW} T. Gao, F. L. Yan, and Z. X. Wang, Quantum secure direct communication by EPR pairs and entanglement swapping, \href{https://doi.org/10.1393/ncb/i2004-10090-1} {Il Nuovo Cimento \textbf{119B}, 313 (2004)}.

\bibitem{GYW} T. Gao, F. L. Yan, and Z. X. Wang, Deterministic secure direct communication using GHZ states and swapping quantum entanglement, \href{https://iopscience.iop.org/article/10.1088/0305-4470/38/25/011} {J. Phys. A \textbf{38}, 5761 (2005)}.

\bibitem{BennettBCJPW} C. H. Bennett, G. Brassard, C. Cr\'{e}peau, R. Jozsa, A. Peres, and W. K. Wootters, Teleporting an unknown quantum state via dual classical and Einstein-Podolsky-Rosen channels, \href{https://journals.aps.org/prl/abstract/10.1103/PhysRevLett.70.1895} {Phys. Rev. Lett. \textbf{70}, 1895 (1993)}.

%11-20
\bibitem{GaoYL} T. Gao, F. L. Yan, and Y. C. Li, Optimal controlled teleportation, \href{https://doi.org/10.1209/0295-5075/84/50001} {Europhys. Lett. \textbf{84}, 50001 (2008)}.

\bibitem{GaoYanW} T. Gao, F. L. Yan, and Z. X. Wang, Controlled quantum teleportation and secure direct communication, \href{https://iopscience.iop.org/article/10.1088/1009-1963/14/5/006} {Chin. Phys. \textbf{14}, 893 (2005)}.

\bibitem{Lo} H. K. Lo and H. F. Chau, Unconditional security of quantum key distribution over arbitrarily long distances, \href{https://science.sciencemag.org/content/283/5410/2050} {Science \textbf{283}, 2050 (1999)}.

\bibitem{Ekert} A. K. Ekert, Quantum cryptography based on Bell's theorem, \href{https://journals.aps.org/prl/abstract/10.1103/PhysRevLett.67.661} {Phys. Rev. Lett. \textbf{67}, 661 (1991)}.

\bibitem{BennettBM} C. H. Bennett, G. Brassard, and N. D. Mermin, Quantum cryptography without Bell's theorem, \href{https://journals.aps.org/prl/abstract/10.1103/PhysRevLett.68.557} {Phys. Rev. Lett. \textbf{68}, 557 (1992)}.

\bibitem{YZhou} Y. H. Zhou, Z. W. Yu, and X. B. Wang, Making the decoy-state measurement-device-independent quantum key distribution practically useful, \href{https://journals.aps.org/pra/abstract/10.1103/PhysRevA.93.042324} {Phys. Rev. A \textbf{93}, 042324 (2016)}.

\bibitem{Bell} J. S. Bell, On the Einstein Podolsky Rosen paradox, \href{https://journals.aps.org/ppf/abstract/10.1103/PhysicsPhysiqueFizika.1.195} {Physics \textbf{1}, 195 (1964)}; On the problem of hidden variables in quantum mechanics, \href{https://journals.aps.org/rmp/abstract/10.1103/RevModPhys.38.447} {Rev. Mod. Phys. \textbf{38}, 447 (1966)}.

\bibitem{Clauser} J. F. Clauser, M. A. Horne, A. Shimony, and R. A. Holt, Proposed experiment to test local hidden-variable theories, \href{https://journals.aps.org/prl/abstract/10.1103/PhysRevLett.23.880} {Phys. Rev. Lett. \textbf{23}, 880 (1969)}.

\bibitem{Freedman} S. J. Freedman and J. F. Clauser, Experimental test of local hidden-variable theories, \href{https://journals.aps.org/prl/abstract/10.1103/PhysRevLett.28.938} {Phys. Rev. Lett. \textbf{28}, 938 (1972)}.

\bibitem{Yan} F. L. Yan, T. Gao, and E. Chitambar, Two local observables are sufficient to characterize maximally entangled states of $N$ qubits, \href{https://journals.aps.org/pra/abstract/10.1103/PhysRevA.83.022319} {Phys. Rev. A \textbf{83}, 022319 (2011)}.

%21-30
\bibitem{Meng} H. X. Meng, J. Zhou, Z. P. Xu, H. Y. Su, T. Gao, F. L. Yan, and J. L. Chen, Hardy's paradox for multisetting high-dimensional systems, \href{https://journals.aps.org/pra/pdf/10.1103/PhysRevA.98.062103} {Phys. Rev. A \textbf{98}, 062103 (2018)}.

\bibitem{Chen} Z. Q. Chen, Bell-Klyshko inequalities to characterize maximally entangled states of $n$ qubits, \href{https://journals.aps.org/prl/abstract/10.1103/PhysRevLett.93.110403} {Phys. Rev. Lett. \textbf{93}, 110403 (2004)}.

\bibitem{DingHYG} D. Ding, Y. Q. He, F. L. Yan, and T. Gao, Quantum nonlocality of generic family of four-qubit entangled pure states, \href{https://doi.org/10.1088/1674-1056/24/7/070301} {Chin. Phys. B \textbf{24}, 070301 (2015)}.

\bibitem{DHYG} D. Ding, Y. Q. He, F. L. Yan, and T. Gao, Entanglement measure and quantum violation of Bell-type inequality, \href{https://doi.org/10.1007/s10773-016-3048-1} {Int. J. Theor. Phys. \textbf{55}, 4231 (2016)}.

\bibitem{BennettDFMRSSW} C. H. Bennett, D. P. DiVincenzo, C. A. Fuchs, T. Mor, E. Rains, P. W. Shor, J. A. Smolin, and W. K. Wootters, Quantum nonlocality without entanglement, \href{https://journals.aps.org/pra/abstract/10.1103/PhysRevA.59.1070} {Phys. Rev. A \textbf{59}, 1070 (1999)}.

\bibitem{Niset} J. Niset and N. J. Cerf, Multipartite nonlocality without entanglement in many dimensions, \href{https://journals.aps.org/pra/abstract/10.1103/PhysRevA.74.052103} {Phys. Rev. A \textbf{74}, 052103 (2006)}.

\bibitem{ZhangZGWO} Z. C. Zhang, K. J. Zhang, F. Gao, Q. Y. Wen, and C. H. Oh, Construction of nonlocal multipartite quantum states, \href{https://journals.aps.org/pra/abstract/10.1103/PhysRevA.95.052344} {Phys. Rev. A \textbf{95}, 052344 (2017)}.

\bibitem{Feng} Y. Feng and Y. Y. Shi, Characterizing locally indistinguishable orthogonal product states, \href{https://doi.org/10.1109/TIT.2009.2018330} {IEEE Trans. Inf. Theory \textbf{55}, 2799 (2009)}.

\bibitem{ZhangGTCW} Z. C. Zhang, F. Gao, G. J. Tian, T. Q. Cao, and Q. Y. Wen, Nonlocality of orthogonal product basis quantum states, \href{https://journals.aps.org/pra/abstract/10.1103/PhysRevA.90.022313} {Phys. Rev. A \textbf{90}, 022313 (2014)}.

\bibitem{WangLZF} Y. L. Wang, M. S. Li, Z. J. Zheng, and S. M. Fei, Nonlocality of orthogonal product-basis quantum states, \href{https://journals.aps.org/pra/abstract/10.1103/PhysRevA.92.032313} {Phys. Rev. A \textbf{92}, 032313 (2015)}.

%31-40
\bibitem{Xu} G. B. Xu, Q. Y. Wen, S. J. Qin, Y. H. Yang, and F. Gao, Quantum nonlocality of multipartite orthogonal product states, \href{https://journals.aps.org/pra/abstract/10.1103/PhysRevA.93.032341} {Phys. Rev. A \textbf{93}, 032341 (2016)}.

\bibitem{WLZF} Y. L. Wang, M. S. Li, Z. J. Zheng, and S. M. Fei, The local indistinguishability of multipartite product states, \href{https://doi.org/10.1007/s11128-016-1477-7} {Quant. Info. Proc. \textbf{16}, 5 (2017)}.

\bibitem{SHalder} S. Halder, Several nonlocal sets of multipartite pure orthognal product states, \href{https://journals.aps.org/pra/abstract/10.1103/PhysRevA.98.022303} {Phys. Rev. A \textbf{98}, 022303 (2018)}.

\bibitem{Jiang} D. H. Jiang and G. B. Xu, Nonlocal sets of orthogonal product states in an arbitrary multipartite quantum system, \href{https://journals.aps.org/pra/abstract/10.1103/PhysRevA.102.032211} {Phys. Rev. A \textbf{102}, 032211 (2020)}.

\bibitem{Rout21} S. Rout, A. G. Maity, A. Mukherjee, S. Halder, and M. Banik, Multiparty orthogonal product states with minimal genuine nonlocality, \href{https://journals.aps.org/pra/abstract/10.1103/PhysRevA.104.052433} {Phys. Rev. A \textbf{104}, 052433 (2021)}.

\bibitem{Li21} M. S. Li, Y. L. Wang, F. Shi, and M. H. Yung, Local distinguishability based genuinely quantum nonlocality without entanglement, \href{https://doi.org/10.1088/1751-8121/ac28cd} {J. Phys. A: Math. Theor. \textbf{54}, 445301 (2021)}.

\bibitem{Terhal} B. M. Terhal, D. P. DiVincenzo, and D. W. Leung, Hiding bits in Bell states, \href{https://journals.aps.org/prl/abstract/10.1103/PhysRevLett.86.5807} {Phys. Rev. Lett. \textbf{86}, 5807 (2001)}.

\bibitem{DiVincenzo} D. P. DiVincenzo, D. W. Leung, and B. M. Terhal, Quantum data hiding, \href{https://ieeexplore.ieee.org/document/985948/} {IEEE Trans. Inf. Theory \textbf{48}, 580 (2002)}.

\bibitem{Eggeling} T. Eggeling and R. F. Werner, Hiding classical data in multipartite quantum states, \href{https://journals.aps.org/prl/abstract/10.1103/PhysRevLett.89.097905} {Phys. Rev. Lett. \textbf{89}, 097905 (2002)}.

\bibitem{Hillery} M. Hillery, V. Buzek, and A. Berthiaume, Quantum secret sharing, \href{https://journals.aps.org/pra/abstract/10.1103/PhysRevA.59.1829} {Phys. Rev. A \textbf{59}, 1829 (1999)}.

\bibitem{Guo} G. P. Guo, C. F. Li, B. S. Shi, J. Li, and G. C. Guo, Quantum key distribution scheme with orthogonal product states, \href{https://journals.aps.org/pra/abstract/10.1103/PhysRevA.64.042301} {Phys. Rev. A \textbf{64}, 042301 (2001)}.

\bibitem{Hsu} L. Y. Hsu and C. M. Li, Quantum secret sharing using product states, \href{https://journals.aps.org/pra/abstract/10.1103/PhysRevA.71.022321} {Phys. Rev. A \textbf{71}, 022321 (2005)}.

%41-50
\bibitem{Markham} D. Markham and B. C. Sanders, Graph states for quantum secret sharing, \href{https://journals.aps.org/pra/abstract/10.1103/PhysRevA.78.042309} {Phys. Rev. A \textbf{78}, 042309 (2008)}.

\bibitem{Rahaman} R. Rahaman and M. G. Parker, Quantum scheme for secret sharing based on local distinguishability, \href{https://journals.aps.org/pra/abstract/10.1103/PhysRevA.91.022330} {Phys. Rev. A \textbf{91}, 022330 (2015)}.

\bibitem{JWang} J. Wang, L. Li, H. Peng, and Y. Yang, Quantum-secret-sharing scheme based on local distinguishability of orthogonal multiqudit entangled states, \href{https://journals.aps.org/pra/abstract/10.1103/PhysRevA.95.022320} {Phys. Rev. A \textbf{95}, 022320 (2017)}.

\bibitem{Halder} S. Halder, M. Banik, S. Agrawal, and S. Bandyopadhyay, Strong quantum nonlocality without entanglement, \href{https://journals.aps.org/prl/abstract/10.1103/PhysRevLett.122.040403} {Phys. Rev. Lett. \textbf{122}, 040403 (2019)}.

\bibitem{Walgate} J. Walgate and L. Hardy, Nonlocality, asymmetry, and distinguishing bipartite states, \href{https://journals.aps.org/prl/abstract/10.1103/PhysRevLett.89.147901} {Phys. Rev. Lett. \textbf{89}, 147901 (2002)}.

\bibitem{Rout} S. Rout, A. G. Maity, A. Mukherjee, S. Halder, and M. Banik, Genuinely nonlocal product bases: Classification and entanglement-assisted discrimination, \href{https://journals.aps.org/pra/abstract/10.1103/PhysRevA.100.032321} {Phys. Rev. A \textbf{100}, 032321 (2019)}.

\bibitem{ZhangZ} Z. C. Zhang and X. D. Zhang, Strong quantum nonlocality in multipartite quantum systems, \href{https://journals.aps.org/pra/abstract/10.1103/PhysRevA.99.062108} {Phys. Rev. A \textbf{99}, 062108 (2019)}.

\bibitem{ShiLHCYWZ} F. Shi, M. S. Li, M. Y. Hu, L. Chen, M. H. Yung, Y. L. Wang, and X. D. Zhang, Strongly nonlocal unextendible product bases do exist, \href{https://doi.org/10.22331/q-2022-01-05-619} {Quantum \textbf{6}, 619 (2022)}.

\bibitem{Yuan} P. Yuan, G. J. Tian, and X. M. Sun, Strong quantum nonlocality without entanglement in multipartite quantum systems, \href{https://journals.aps.org/pra/abstract/10.1103/PhysRevA.102.042228} {Phys. Rev. A \textbf{102}, 042228 (2020)}.

\bibitem{ZGY} H. Q. Zhou, T. Gao, and F. L. Yan, On orthogonal product stes with strong quantum nonlocality on plane structure,  \href{https://journals.aps.org/pra/abstract/10.1103/PhysRevA.106.052209} {Phys. Rev. A \textbf{106}, 052209 (2022)}.


%51-60
\bibitem{Che} B. C. Che, Z. Dou, M. Lei, and Y. X. Yang, Strong nonlocal sets of UPB, \href{https://arxiv.org/abs/2106.08699} {arXiv:2106.08699v2}.

\bibitem{ShiYCZ} F. Shi, Z. Ye, L. Chen, and X. D. Zhang, Strong quantum nonlocality in $N$-partite systems, \href{https://journals.aps.org/pra/abstract/10.1103/PhysRevA.105. 022209} {Phys. Rev. A \textbf{105}, 022209 (2022)}.

\bibitem{He} Y. Y. He, F. Shi, and X. D. Zhang, Strong quantum nonlocality and unextendibility without entanglement in $N$-partite
systems with odd $N$, \href{https://arxiv.org/abs/2203.14503} {arXiv:2203.14503v1}.

\bibitem{Cohen08} S. M. Cohen, Understanding entanglement as resource: Locally distinguishing unextendible product bases, \href{https://journals.aps.org/pra/abstract/10.1103/PhysRevA.77.012304} {Phys. Rev. A \textbf{77}, 012304 (2008)}.

\bibitem{Bandyopadhyay18} S. Bandyopadhyay, S. Halder, and M. Nathanson, Optimal resource states for local state discrimination, \href{https://journals.aps.org/pra/abstract/10.1103/PhysRevA.97.022314} {Phys. Rev. A \textbf{97}, 022314 (2018)}.

\bibitem{ZhangSSGQW} Z. C. Zhang, Y. Q. Song, T. T. Song, F. Gao, S. J. Qin, and Q. Y. Wen, Local distinguishability of orthogonal quantum states with multiple copies of $2\otimes2$ maximally entangled states, \href{https://journals.aps.org/pra/abstract/10.1103/PhysRevA.97.022334} {Phys. Rev. A \textbf{97}, 022334 (2018)}.

\bibitem{LiGZW} L. J. Li, F. Gao, Z. C. Zhang, and Q. Y. Wen, Local distinguishability of orthogonal quantum states with no more than one ebit of entanglement, \href{https://journals.aps.org/pra/abstract/10.1103/PhysRevA.99.012343} {Phys. Rev. A \textbf{99}, 012343 (2019)}.

\bibitem{ZhangWZ} Z. C. Zhang, X. Wu, and X. D. Zhang, Locally distinguishing unextendible product bases by using entanglement efficiently, \href{https://journals.aps.org/pra/abstract/10.1103/PhysRevA.101.022306} {Phys. Rev. A \textbf{101}, 022306 (2020)}.

\bibitem{Bandyopadhyay16} S. Bandyopadhyay, S. Halder, and M. Nathanson, Entanglement as a resource for local state discrimination in multipartite systems, \href{https://journals.aps.org/pra/abstract/10.1103/PhysRevA.94.022311} {Phys. Rev. A \textbf{94}, 022311 (2016)}.

\bibitem{Nielsen} M. A. Nielsen and I. L. Chuang, \textit{Quantum Computation and Quantum Information} (Cambridge University Press, Cambridge, 2000).


\end{thebibliography}
\end{document}